\documentclass{aa}
\usepackage{times,graphicx}
\usepackage{ulem,widetext}

\usepackage{mathtools}
\begin{document}

\title{Collisional   effects in modeling solar polarized lines}
\author{ M. Derouich\thanks{derouichmoncef@gmail.com, aldarwish@kau.edu.sa} and S. Qutub}

\institute{ 
Astronomy and Space Science Department, Faculty of Science, King Abdulaziz University, P.O. Box 80203, Jeddah 21589, Saudi Arabia   }

\titlerunning{Collisional    effects on the solar polarization}
\authorrunning{}

\date{Received 4 February 2023 / Accepted 10 December 2023 }
\abstract
 {Rigorous implementation of the effects of collisions in modeling the formation of the polarized solar lines is of utmost
importance in order to realistically analyze the available, highly sensitive solar spectropolarimetric observations. Indeed, even when
an observation seems to fit well with theory, one can misinterpret results if important effects due to collisions are not correctly
implemented in the modeling process.}
{  
We point out inconsistencies in the models adopted to implement the Paschen Back effect together with collisional effects on the
solar linear polarization formed by scattering of anisotropic radiation. Because the significance of these inconsistencies increases as
polarization becomes increasingly responsive to collisions, we investigate the range of hydrogen densities $n_\text{H}$ to which the polarization
is sensitive. 
}
{  
We used the density matrix formalism in the tensorial irreducible basis, which was developed within the theory of atom-radiation interaction and of atomic collisions. We solved the statistical equilibrium equations for multi-level atoms with hyperfine
structure (HFS) in order to evaluate the collisional depolarization of levels of the D1-D2 lines of the K I atom.}
{
{  We find  that collisions play a prominent role, particularly at hydrogen densities of between  10$^{13}$ and 10$^{16}$  cm$^{-3}$}.  
 }
{So far, analyses of polarized lines formed in the presence of solar magnetic field have incorporated, if at all, collisional
rates calculated assuming zero magnetic field. This could be a good approximation in the Hanle regime but not in the Paschen Back
regime. For typical quiet Sun magnetic fields, the latter regime could be reached, and level-crossing takes place in several atomic
systems. Therefore, one must be careful when using collisional rates calculated in the zero-field case to interpret linear polarization
formed in magnetized media. }
\maketitle
\section{Introduction} \label{zero-magnetic field case}
In the solar photosphere,  atoms and molecules that emit polarized light are exposed to collisions with neutral hydrogen atoms. The effects of these collisions must be correctly incorporated into the model of the line formation in order to properly interpret the observed polarized lines.  The internal states of the perturbed atom can be described by the density matrix components expressed in the   tensorial basis,  $\rho_q^k$. We adopt the usual notations, where $k$ denotes   
the tensorial rank (order) and $q$ quantifies the coherence  between  the levels.      Calculating   the polarization of spectral lines requires the determination of the    elements $\rho_q^k$. Theoretically, collisions produce  distinct but correlated effects on the   $\rho_q^k$  values  and consequently    lead to variation of    the   polarization of the emitted light.  In this context, the following factors must be taken into account:
\begin{itemize}  
\item  
  Isotropic  collisions produce   gain terms due to transfer rates and loss terms due to   relaxation    rates. Collisional rates must be calculated in the tensorial basis to quantify the contribution  of these collisions to the statistical equilibrium equations (SEEs)  giving  the variation of    $\rho_q^k$ values (e.g. Derouich et al. 2003, Sahal-Br\'echot et al. 2007, Derouich 2020). 
\item 
For a given radiative transition, collisions cause  perturbation to the energies of the levels participating in the transition, which results in broadening of the corresponding line; the rate of this broadening is denoted  $w$. The value of $w$ depends on the density of the perturbers and the Hamiltonian of the system,  as well as the dynamics of the collisions  (e.g., Derouich et al. 2015). We note that $w=\frac{\gamma_E}{2}$, where $\gamma_E$  is the rate of elastic collisions, which usually  enters the branching ratios
of the redistribution matrices (more details about $w$ and $\gamma_E$ are available in Derouich et al. 2015).
\item  
Modeling the polarization profiles,  particularly if partial frequency redistribution (PRD) effects must be considered, requires the correct evaluation of  the  role of collisions   (e.g., Nagendra et al. 2020).
\end{itemize}  
More details about collisions and their effects on polarized lines can be found in our previous work published over the last 20 years (see Derouich 2020 and references therein).
 
In the solar conditions, collisional rate calculations available in the literature    are performed for a Hamiltonian $H_0,$ which  completely neglects the impact of a magnetic field during the collisions.  In such calculations, the basis used consists of  the vectors $|  \alpha J M_J \rangle,$   which are the  eigenvectors of $H_0$;  the total angular momentum, $J$, is considered to be well defined and to constitute a good quantum number  (here $M_J$ is the projection of $J$ along the quantization axis and $\alpha$   summarizes the
electronic configuration quantum numbers). Similarly, 
if the problem necessitates the inclusion of HFS,   the total angular momentum $F$ is taken to be a good quantum number and the basis $\{|  \beta F M_F \rangle\}$   becomes  the eigenbasis of $H_0$ (here $M_F$ represents the projection of $F$ and $\beta$ denotes the other quantum  numbers associated to the level).    In our investigation, we  take into account the   coherence between the different   $F$-levels within a given $J$-level.      

\section{ Challenges of modeling polarized lines in strong magnetic fields and collisions}
 Strictly speaking, one cannot use     elastic collisional broadening  rates $w$ and depolarization   rates $D^k$  calculated in a zero-magnetic field case to model  polarized line formation in the presence of  a  nonzero  magnetic field.  This strict condition might be supplemented with a looser one stating that zero-field collisional rates  can be adopted in modeling cases where the magnetic field is sufficiently weak,     specifically within the Hanle regime, where the angular momenta $F$ and $J$ are   good  (well-defined) quantum numbers.

  Both collisional line broadening and collisional depolarization share similar approximations, and for a given level they are calculated using the same scattering collisional matrix $S$ (Derouich et al. 2003, Kerkeni et al. 2004, Derouich et al. 2015, Sahal-Br\'echot \& Bommier (2014, 2019)). The main distinction is that $D^k$ pertains to a single state, whereas w involves two states. However, this does not alter the fact that the elastic collision contribution to line broadening ($w$) and (de)polarization ($D^k$ rates) of the lower and upper levels are expressed in terms of the same $S$-matrix elements derived from calculating the interaction potential and solving the Schr\"odinger equation. The scattering collisional matrix $S$ strongly depends on the Hamiltonian of the system and its eigenbasis.  As a result, both $D^k$  and $w$ depend on the presence of a magnetic field. 

The similarities  between $D^k$  and $w$  have been well documented  and are firmly established. For instance, Derouich et al. (2003) used the Anstee-O'Mara-Barklem (ABO) theory (as described by Barklem \& O'Mara  1998) to calculate $w$, and subsequently employed the same interaction potential and solved the same Schr\"{o}dinger equation to determine the depolarization rates $D^k$.  
It is important to emphasize that the broadening $w$,  {as well as $D^k$}, are known to be dependent on the $J$-level (in cases involving only fine structure) or $F$-level (when considering hyperfine structure) (see, e.g., {Nienhuis   1976, Omont  1977}, Green 1988,  Belli et al.  2000, Kerkeni et al  2004,  Buffa \&   Tarrini  2011a,b,  Sahal-Br\'echot  \& Bommier 2014, 2019, {Derouich 2020}).

     A sufficiently strong magnetic field (as in the case of the incomplete Paschen-Back  (PB) regime) greatly modifies the structure of the energy levels and induces level crossings and anti-crossings. Therefore, quantum numbers such as $J$ and $F$, which are  good quantum numbers  in the absence of a magnetic field, lose their physical sense in the PB regime. {Consequently, zero-field elastic collisional rates, such as the broadening $w(JF)$ and the depolarization rate $D^k(JF),$ lose their physical sense}.   Diagonalization of the Hamiltonian is necessary, and an eigenbasis and eigenvalues with new good quantum numbers must be obtained for each magnetic field strength.  
     In this sense,    previous and ongoing  works including  the effects of collisions   in the presence of   arbitrary magnetic fields appear  to be inconsistent   (see, e.g., Kerkeni \& Bommier 2002,   {Bommier (2017, 2018)}, Nagendra et al.   2020,  Alsina Ballester et al. 2021, Alsina Ballester  2022).

  Kerkeni \& Bommier (2002)  investigated the ffects of mixing zero-field depolarization collisional rates with arbitrary (strong) magnetic field in the case of the Na I D1 and D2 lines where the   {PB effect (also called the Back-Goudsmit  effect)}    is reached. In addition  to the depolarization rates, these authors included the  collisional broadening rates  calculated in the zero-field case.   As shown  in Figure 1,
  the PB effect    induces crossings among the hyperfine sublevels of the $^2P_{3/2}$ state of Na I  at a magnetic field strength of approximately 15 Gauss. This value can be reached in the lower chromosphere   where the D-Na I lines are formed. 
  For instance,   collision effects at a density of $10^{16}$ cm$^{-3}$  are mixed together in the same model with magnetic field going from 0.1 to 100 Gauss, which   severely  influences their conclusions  (see  Figure 8 of Kerkeni \& Bommier  2002).
In this sense, these latter authors pointed out that the loops visible in their polarization--magnetic fields diagrams (Figs 5--7 of Kerkeni \& Bommier 2002)   appear  and disappear due to strong magnetic and collisional effects.     This implies  that these authors, in their interpretation, 
inconsistently take   into   account   both effects happening in a strong magnetic field (incomplete PB regime)  and collisional effects calculated in a zero-magnetic-field case.   It is worth mentioning  that the   PB regime is reached in the case of the D2 upper level   $^2P_{3/2}$   for lower magnetic field values than those needed to reach the  {PB}  regime in the  D1 upper level   $^2P_{1/2}$. However,   the atomic polarization degrees  of the  HFS   levels within the  $^2P_{1/2}$ state  are   inaccurately and inconsistently determined in the {PB regime,} because they are strongly coupled to those within the  $^2P_{3/2}$ state  by the SEE.

The redistribution function    obtained  by Bommier (2017, 2018) incorporates the zero-field collisional broadening rate $w$ (denoted $\Gamma_E$ in Bommier 2017) within the {PB} regime.  Bommier (2017, 2018) did not acknowledge the inherent inconsistency of including elastic collisional effects calculated in a zero-field scenario within the framework of the {PB} regime.\footnote{The omission of $D^k$ in Bommier  (2017, 2018) does not indicate that she recognized the inconsistency of utilizing zero-field rates in the presence of strong magnetic fields. Instead, she excluded these rates  because of difficulties related to factorization problems, which prevent  an analytical solution of the SEE for any B $\ne$ 0  {(we refer readers to the last paragraph of section 2.2.1 of Bommier 2017)}.}     Consequently, works that based their calculations on the methodology presented by Bommier (2017, 2018) exhibit inconsistencies. As can be seen in Alsina Ballester (2022), the   inconsistency   should be {encountered} in the case of the   K I  atom, because the {incomplete}  {PB} regime is reached at a magnetic field of B $\sim$ 5 Gauss. Furthermore, interpretation of the polarization of the K I lines for  magnetic fields with B $>$ 5 Gauss  faces a serious problem because of collisional effects, which are not  known  in the {incomplete}  {PB} regime.  Sowmya et al. (2015, 2019) examined Li I D1 and D2 lines   within a collisionless model   and  encountered the incomplete {PB} regime at magnetic fields of as low as around 2 Gauss.  If one were to consider the Li atoms in the incomplete {PB} regime and use zero-field collision rates, a similar inconsistency would be   encountered.
 
\begin{figure}[h]
\begin{center}
 \includegraphics[width=7.5 cm]{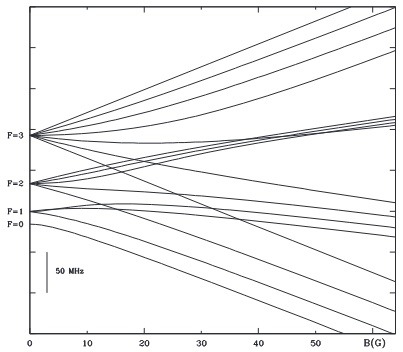}  
 \caption{Crossings between  the hyperfine sublevels of the state $^2P_{3/2}$ of Na I
due to  magnetic-field-strength effects (this figure is taken from Landi Degl'Innocenti \& Landolfi 2004)}
\label{Fig-1-Na-New-paper}  
\end{center}
\end{figure}

 There are two essential steps in the computation of the collisional depolarization or broadening rates:
 \begin{enumerate} 
\item  Determination of the Hamiltonian of the system.  We note that, in the PB regime, the magnetic Hamiltonian cannot be safely neglected    when one calculates the total Hamiltonian of the system. 
\item Treatment of 
 the collision dynamics which needs to be performed in a suitable  eigenbasis of the Hamiltonian determined for each value of the magnetic field. 
 \end{enumerate} 
  Usually, one determines the collisional rates as a function of the temperature after averaging the cross-sections over the velocity distribution function and multiplying by $n_\text{H}$ (e.g., Derouich et al. 2003).  Typically, collisional rates vary with temperature as $\sim T^{0.4}$. However, in the future, significant effort should be devoted to the determination of the rates as a two-variable function  $D^k(T,B)$ and $w(T,B),$ which is a challenging problem. 

  \section{Collisional effects in a multilevel atom with hyperfine structure }

\begin{figure}[h]
\begin{center}
 \includegraphics[width=7 cm]{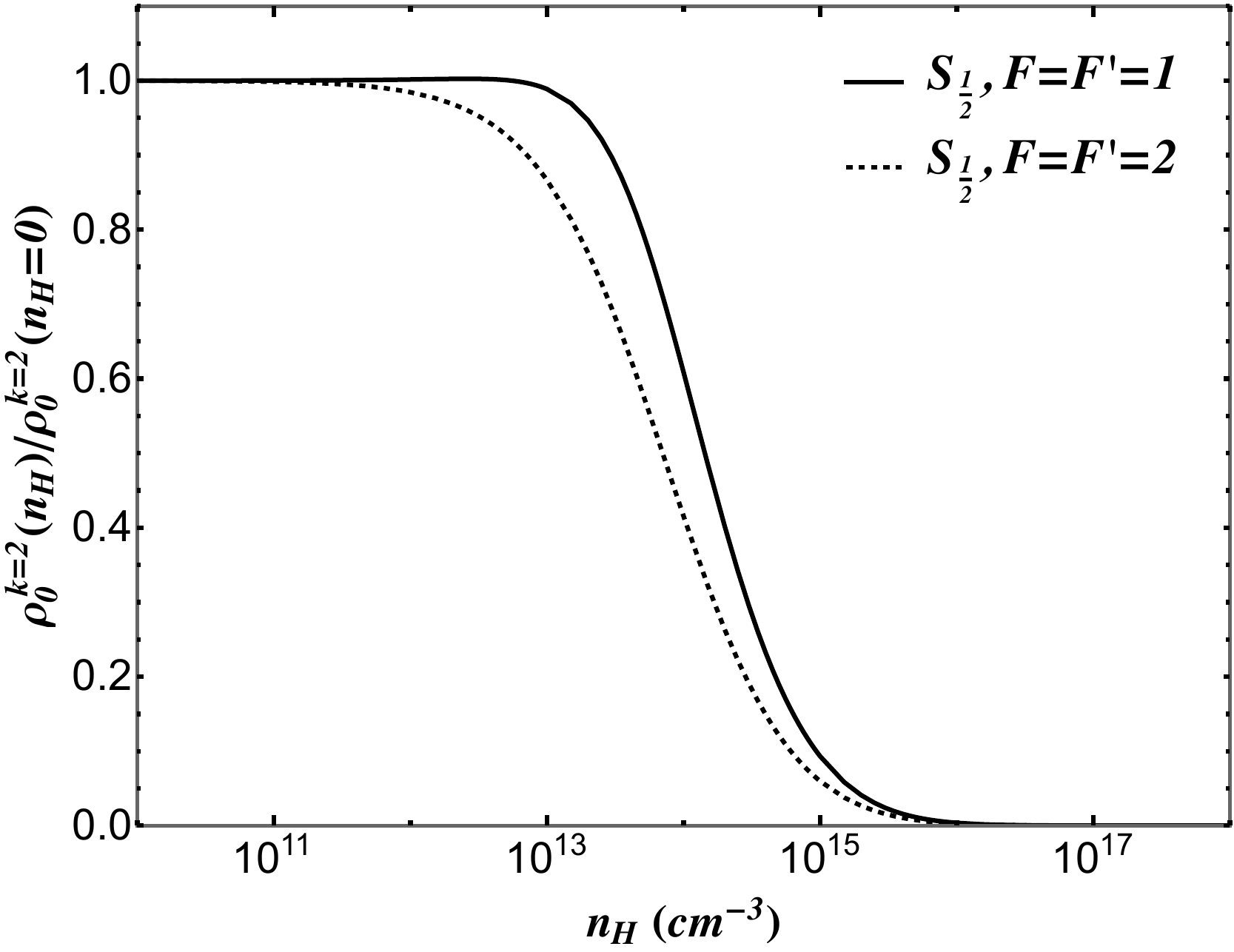}  
  \includegraphics[width=7  cm]{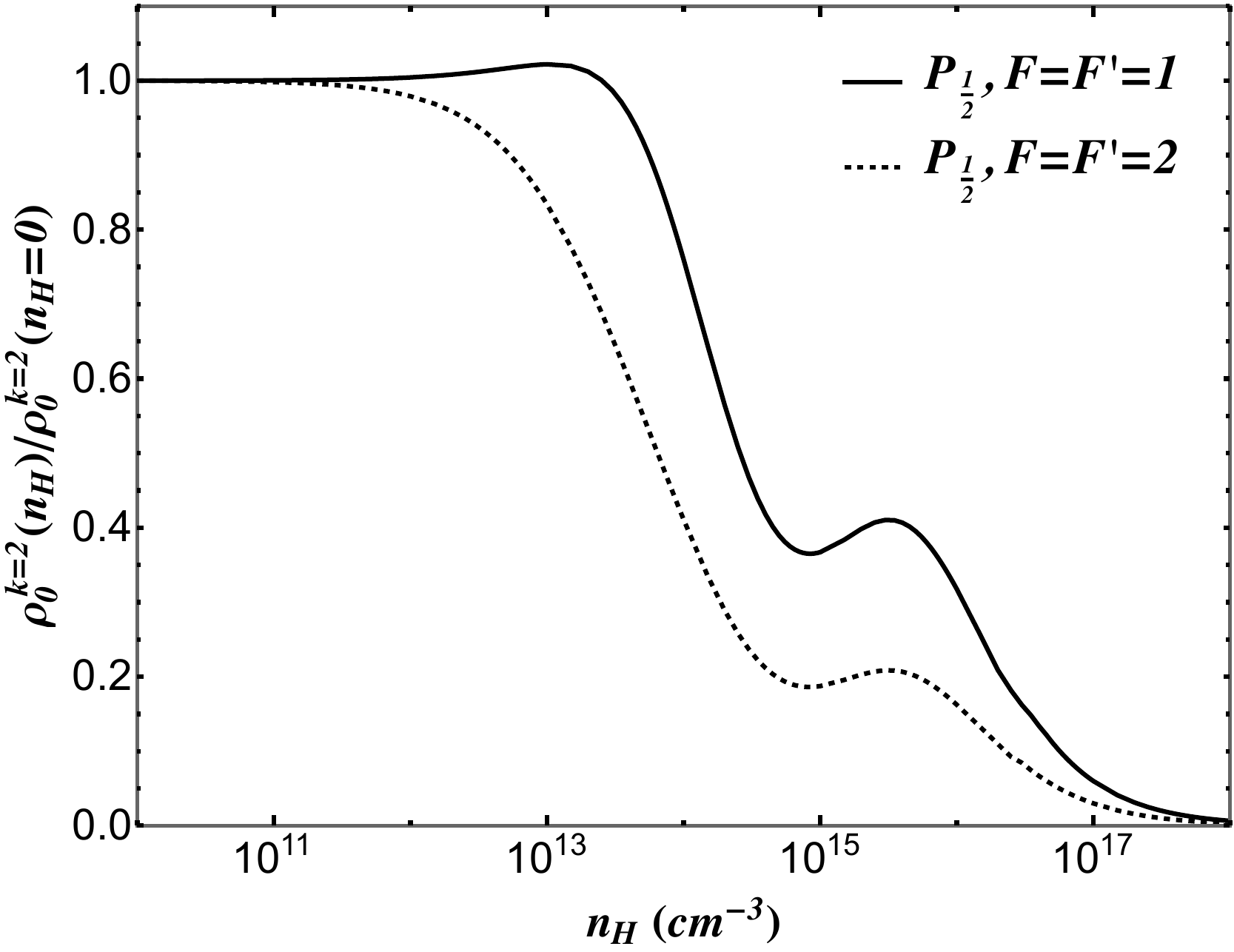}
    \includegraphics[width=7  cm]{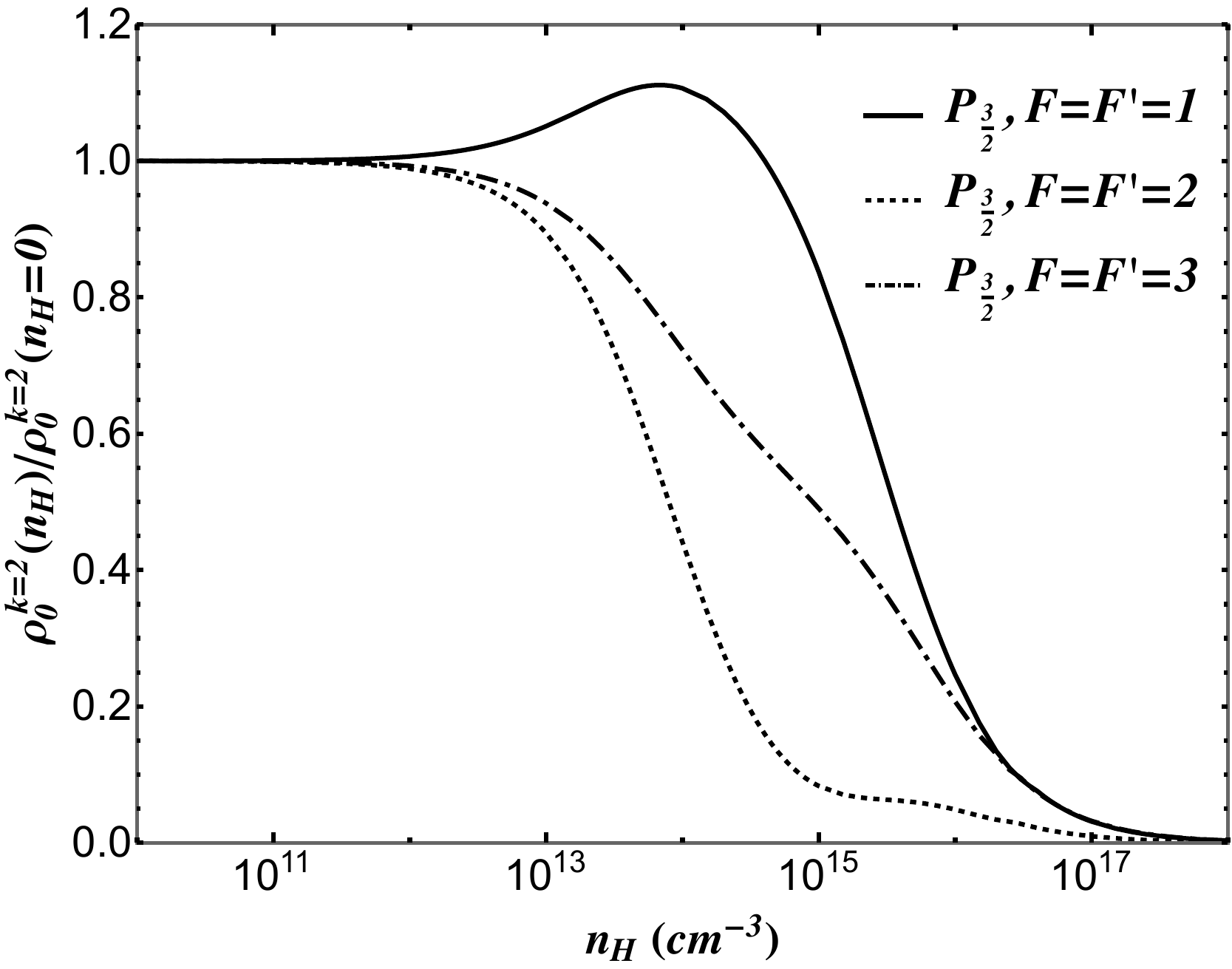}
 \caption{Emergent alignment for different  HFS  KI  levels as a function of the density of neutral hydrogen $n_\text{H}$. }
\label{Fig-1-Na-New-paper}  
\end{center}
\end{figure}

As explained above, the inclusion of collision rates calculated in a zero-magnetic-field case  in modeling the formation of polarization  in the  {PB} regime is conceptually wrong. The severity   of the effects of this error depends  on how much the polarization is sensitive to collisions. 

To evaluate the (de)polarizing effect of isotropic collisions, at first glance, one might compare the  depolarization rates $D^k$    to the appropriate radiative rates  associated with the transitions between the  HFS   levels.  However, a more precise approach is to solve the SEE for the case of a multilevel atom with hyperfine structure.

 We adopt the case of the  multilevel atom with hyperfine structure detailed in Landi Degl'Innocenti \& Landolfi (2004) by incorporating the influence of coherences between distinct $F$-levels within a given $J$-level. We include zero-field collisional rates in our analysis, focusing solely on the effects of collisions and radiative rates, while disregarding the magnetic field. 
 We solved  the SEE for the   HFS   levels involved in  modeling the formation of the D1 and D2 lines of the K I atom.

 We developed  a numerical code that 
  takes data describing HFS quantum levels    as input,     together with 
  the number of photons per mode $\bar{n}$   and the corresponding anisotropy factor $w_a$.  {Einstein coefficients required in the SEE are extracted   from the NIST database and, using the results of   Manso Sainz \& Landi Degl’Innocenti (2002), we determined that $\bar{n} $(D1)=0.001906, $w_a$(D1)=0.08365,  $\bar{n}$(D2)=0.001875,  and  $w_a $(D2)=0.0842}.     In addition  {to the expressions of the radiative rates taken from Landi Degl'nnocenti \& Landolfi (2004)}, the numerical code incorporates  collisional data  {as input} to compute the evolution of the atomic density matrix. {Its evolution}  is affected by the gain terms, which are called  collisional polarization transfer rates and are denoted $D^k(J'FF'  \to JFF')$, and  {by}   the loss terms associated with relaxation or depolarization rates, denoted $D^k(JFF')$.   
The   $D^k(JFF')$ and $D^k(JFF'  \to J'FF')$ rates are obtained through a methodology first proposed by  Nienhuis  (1976) and Omont (1977), 
  where $D^k(JFF')$ and $D^k(J'FF'  \to JFF')$ are represented as   linear combinations of the $D^k(J)$ and $D^k(J' \to J)$ rates, with these latter taken from Derouich (2012; see also Derouich 2020).   To conveniently  investigate the effect of collisions on polarization, it is essential to simultaneously take into account gain and loss collisional effects. We recognize that our current implementation in the SEE does not encompass all possible collisional rates governing coherence gain or loss.  Nevertheless, our present inclusion of these aspects is novel and adequately conveys the core message of our paper. The incorporation of all possible rates poses a numerical challenge that we plan to address in the future.

    We consider a slab     in the solar atmosphere  containing  K I atoms and illuminated 
anisotropically by  photospheric unpolarized radiation field.  
Assuming that the incident radiation has cylindrical
symmetry around the local solar vertical,  at a frequency $\nu$  only the  radiation matrix elements  $J_0^{k_r=2}(\nu)$ with   $k_r =$ 0  and $k_r = 2$ 
 are needed to describe the incident radiation where $\bar{n}$= $J^0_0$ ($c^2$ /2 $h$ $\nu^3$)  and  $w_a=$ $\sqrt{2}$ $\frac{J^2_0}{J^0_0}$  (see, e.g., Trujillo Bueno 2001, Landi Degl'Innocenti \& Landolfi 2004, Derouich 2008).     
 In these conditions, due to the cylindrical symmetry of the problem, only the linear polarization    is formed by scattering.

The D1 and D2  K I lines  result from the atomic transitions $^2S_{1/2}$--$^2P_{1/2}$   and  $^2S_{1/2}$--$^2P_{3/2}$, respectively. The  K I atom has nuclear spin $I = 3/2$. Therefore, each of the levels $^2S_{J=1/2}$ and $^2P_{J=1/2}$     splits into two 
  HFS   levels $F = 1$ and $F = 2$. The D2 upper level $^2P_{3/2}$ has $J=3/2$ and therefore splits into four   HFS   levels, $F=$ 0, 1, 2, and 3. Therefore,
  eight   HFS   levels are involved in the atomic model   considered to simulate  the formation of the K I D lines.  By taking into account only the even $k$-orders,  with $q=0$ give that the coherence is neglected, the SEE have 36 unknowns representing the density matrix elements $\rho_0^{k}(L_J, FF')$ with even $k$ values.  We note that    the tensorial order $k$  varies  from  $|F-F'|$ to  $F+F'$,  implying that, for example, the maximum value of $k$ is $k_{max}$ = 6 for $F=F'=$ 3. In our calculation, we take into account all orders $k$.
  We recall    that $D^k$  rates  are $q$-independent because the  collisions are considered to be isotropic.

  We compute the density matrix elements of all $F$-levels for the case of a tangential observation in a  slab in the solar    atmosphere. 
Our code provides   all values of density matrix elements associated to the K I   HFS   levels as output.   Analytical expressions of the SEE are    
 given in the Appendix A   to ensure that they are easily accessible to all readers.

According to Bruls et al. (1992) and   Alsina Ballester (2022),  the   line-core  profile of the K I D lines    should be formed at photospheric heights corresponding to hydrogen densities     $n_\text{H}$ ranging between  $\sim$ $10^{14}$ and $10^{16}$ cm$^{-3}$.    In the present work, we   conduct   a comprehensive analysis by scanning a range of hydrogen densities $n_\text{H}$ going from $10^{12}$ cm$^{-3}$ to  $10^{18}$ cm$^{-3}$  
  in order to fully investigate the possible role of collisions in the formation of the   K I D lines. We  adopt a photospheric temperature of   $T=$5780 K. 
 Fig. 2 shows the ratio $\frac{\rho_0^{k=2}(n_H)}{\rho_0^{k=2}(n_H=0)}$ , which gives the variation of the alignment as a function of $n_\text{H}$. It can be   seen that collisions begin to 
 influence the alignment of the HFS 
 levels associated to the $^2S_{1/2}$ for  $n_\text{H}$ $\sim$ $10^{13}-10^{14}$ cm$^{-3}$. The ground-level $^2S_{1/2}$ is almost completely depolarized for  $n_\text{H}$ $>$ $10^{15}$ cm$^{-3}$. 
  The density matrix element $\rho_0^{k=2} (P_\frac{3}{2}, 3 \, 3)$, which reflects the alignment of hyperfine level $F=3$ within $^2P_{3/2}$, has two remarkable features. First, it exhibits a wide range of variation covering four orders of magnitude from $10^{13}$ to $10^{17}$ cm$^{-3}$. Second,  it is noteworthy that this effect becomes evident at relatively low values, commencing at just $10^{13}$ cm$^{-3}$. 
 
   To understand this behavior,  in the absence of collisions, we  determined an analytical expression giving the density matrix elements  $\rho_0^{k=2} (^2P_{3/2}, 3 \, 3)$ and we find that it  is expressed  as a function of the density matrix elements $\rho_0^{k} (^2S_{1/2}, 2 \, 2)$:
     \begin{eqnarray}  \label{eq_1}
 \rho_0^{k=2} (P_\frac{3}{2}, 3 \, 3) &=& \frac{\sqrt{\frac{21}{10}} \; B  \; J_0^{k_r=2}(\nu_{{P_\frac{3}{2},S_\frac{1}{2}}})}{5 \; A}
\end{eqnarray}
$$ \nonumber \\  \times {\  {\rho_0}}^{ {k}= {0}} ( {S}_\frac{ {1}}{ {2}},  2 \, 2)  	 +   \\ \bigg{[ } \frac{\sqrt6\ \; B  \; J_0^{k_r=0}(\nu_{{P_\frac{3}{2},S_\frac{1}{2}}})}{5 \; A  } \\  + $$ $$ \frac{2\sqrt3\ B \; J_0^{k_r=2}(\nu_{{P_\frac{3}{2},S_\frac{1}{2}}})}{35  \; A  }\bigg{] } \times {\rho}^{ {k}= {2}}  ({S}_\frac{ {1}}{ {2}},  2 \, 2) $$ $$
 + \frac{\sqrt{\frac{3}{5}} \; B \; J_0^{k_r=2}(\nu_{{P_\frac{3}{2},S_\frac{1}{2}}})}{35 \; A } \times {\rho}^{{k}={4}}({S}_\frac{1}{2},   2 \, 2) 
$$
where $B$ and $A$    refer to the Einstein coefficients associated with the $(S_\frac{1}{2} \rightarrow  P_\frac{3}{2})$  transition.
Equation  (\ref{eq_1}) can be written numerically as:
$$ \rho_0^{k=2} (P_\frac{3}{2}, 3 \, 3)  = 6.47  \; 10^{-5} \times {\  {\rho_0}}^{ {k}= {0}} ( {S}_\frac{ {1}}{ {2}}, 2 \, 2)  $$
 $$+   1.86 \; 10^{-3} \times {\rho}^{ {k}= {2}}  ({S}_\frac{ {1}}{ {2}}, 2 \, 2) $$
\begin{equation}  
{\centering + 4.94  \;  10^{-6} \times {\rho}^{{k}={4}}({S}_\frac{1}{2}, 2 \, 2)  } 
   \label{eq_2}
\end{equation}

  Equations (\ref{eq_1})  and (\ref{eq_2}) show  that the alignment of level $(^2P_{3/2},  3 \, 3)$ is connected only to the density matrix elements of level $(^2S_{1/2},  2 \, 2)$.   
   
 We find that the values of $\rho^{k=2,4}(^2S_{1/2}, 2 \, 2)$  begin to exhibit a noticeable decrease when the hydrogen density $n_\text{H}$ is approximately $10^{13}$ cm$^{-3}$, and they become zero at around $n_\textrm{H} \approx 10^{15}$ cm$^{-3}$.   Therefore, for $n_\textrm{H} > 10^{15}$ cm$^{-3}$, the only remaining density matrix element capable of providing alignment to the level  $(^2P_{3/2}, 3 \, 3)$ is the population  $\rho^{k=0}(^2S_{1/2}, 2 \, 2)$.  The continued existence of  $\rho^{k=0}(^2S_{1/2}, 2 \, 2)$, and therefore the survival of  $\rho^{k=2}(^2P_{3/2}, 3 \, 3)$,  relies on the gain in population from $(^2S_{1/2}, 1 \, 1)$ by collisional transfer rates, which results in a population increase counterbalancing the relaxation process. If we neglect the transfer rates and focus  on  the relaxation rates, the population $\rho^{k=0}(^2S_{1/2}, 2 \, 2)$, and therefore $\rho^{k=2}(^2P_{3/2}, 3 \, 3)$,  would be completely eliminated at around $n_\textrm{H}$=$10^{15}$ cm$^{-3}$. Conversely, when we take population transfer (gain) into account,  although  $\rho^{k=2}(^2P_{3/2}, 3 \, 3)$   begins to   decrease at $n_\textrm{H} \approx 10^{13}$ cm$^{-3}$    due to the decrease in  $\rho^{k=2,4}(^2S_{1/2}, 2 \, 2)$,  it keeps decreasing but  does not reach zero until $n_\textrm{H} = 10^{17}$ cm$^{-3}$,   as illustrated in Fig.  2.   
 
On the other hand, an analogous analysis of the SEE can provide insight into the behavior  of other $F$-states within the $P$-state.  Specifically, it elucidates the similar decrease in both $^2P_{1/2}$ and $^2S_{1/2}$ as the hydrogen density $n_\text{H}$ increases, spanning a range from  $10^{13}$ to $10^{15}$ cm$^{-3}$. It is worth mentioning that,   similarly,  Trujillo Bueno et al. (2002) noted  an equal sensitivity of the $^2P_{1/2}$ and $^2S_{1/2}$ HFS levels to the depolarizing influence of a magnetic field.  That being said, it should be noted that the alignments of $F$-levels related to $S$-state  are definitively destroyed by collisions only when $n_\text{H}$ exceeds $10^{15}$ cm$^{-3}$. However,  the alignments of  $F$-levels associated to the $^2P_{1/2}$   are clearly destroyed  by collisions only for 
$n_\text{H}$ $>$ $10^{17}$ cm$^{-3}$. We also note that,  for the $F$-levels of the $^2P_{1/2}$ state, a minor increase occurs at approximately  $n_H \simeq 10^{16}$ cm$^{-3}$ (see Fig. 2).   It is    noteworthy that the enhancement of   alignment by isotropic collisions within a specific range of hydrogen density    is not a novel result. This effect was observed by  Derouich et al. 2007, even when HFS was entirely disregarded (see Derouich et al. 2007).  By analogy, in a specific range of magnetic values, a localized enhancement of the alignment of the $^2P_{3/2}$ ($F=1$ and $F=2$) levels  has also been found; see Trujillo Bueno et al. (2002).  By taking into account collisional effects on all   HFS   levels, we conclude that the polarization of the K I D lines   is sensitive  to the presence of  isotropic collisions for a large range of $n_H,$ going from $\sim$ $10^{13}$ cm$^{-3}$ to $\sim$ $10^{17}$ cm$^{-3}$.\footnote{We note that the behavior of the   $\rho_0^{k=2}$ when depolarizing collisions are increased via increasing $n_\text{H}$ is similar to that obtained when the depolarizing effect of a  magnetic field is increased. However, when the magnetic field is increased, the alignment is not completely destroyed and a residual remains, but when collisions increase ($n_\text{H}$ $\ge$ $10^{17}$) the alignment becomes zero ({there is} no saturation regime for collisions).}

D1- and D2-type lines of alkali atoms have the same atomic model and their collisional rates with atomic hydrogen (see Derouich 2012) ---as well as their Einstein coefficients--- are of the same order of magnitude as those of the K I atomic system. Therefore, like the K I D lines, other alkali atoms should be very sensitive to collisions for photospheric hydrogen densities, which complicates their interpretation, especially in the {PB} regime. 

 \section{Important remark}
We note that, in addition to the inconsistency arising from the use of zero-magnetic rates in conditions where polarization is generated in the presence of a strong magnetic field, there is also a misapplication of collision rates   in the absence of a magnetic field.  To  model  the collisional effect on the $Q/I$ profiles   of  K I D lines, Alsina Ballester (2022) used  the  destruction of the orientation    rates $D^{k=1}$,  obtained  according to the 
formulae given in   Section (7.13) of Landi Degl'Innocenti \& Landolfi (2004). However, it is important to note that   $D^{k=1}$ rates are only relevant in   studies    where the scattering circular polarization   exists and is coupled to linear polarization.\footnote{   Circular polarization can be produced, for example, by illuminating the atom with circularly polarized radiation, or by the   alignment-to-orientation conversion mechanism, which can be triggered by the PB effect (see, e.g., Landi Degl'Innocenti \& Landolfi 2004).}  In this sense, the inclusion of $D^{k=1}$ rates by Alsina Ballester (2022) in the  zero-field case does not seem to be  appropriate.    On the other hand, the rates due to elastic collisions $\gamma_E$  should also
be of importance  and can be found by applying the analytical expressions provided by   Derouich et al. (2015), where a numerical  model unifying $\gamma_E$ (or  broadening $w$) and collisional depolarization
  rates  is obtained by using accurate genetic programming (GP) numerical methods. We note that several works (e.g., Stenflo  1994, Faurobert et al. 1995, Berdyugina \&   Fluri 2004, Smitha et al. 2014, Alsina Ballester 2022),  in the framework of the solar application, assume    simple
  relations between broadening and   depolarizing   rates, which  is difficult to justify    either theoretically or numerically.  For instance, for atomic and molecular collisions, there is no justification for the widely used relation  $\frac{D^2}{\gamma_E}=0.5$ (or 0.38 or 0.1) (e.g.,  Faurobert et al. 1995, Berdyugina \&   Fluri 2004, Smitha et al. 2014). 
  \section{Conclusions}
  Current collisional data dedicated to solar applications are calculated in the absence of external magnetic field. Such data are useful only for polarimetric diagnostics in unmagnetized plasmas or in media with sufficiently weak magnetic fields, such as in the Hanle effect regime. 
     Incorporating collision rates derived in a zero magnetic field context into models for polarization formation within the incomplete or complete   PB regime is conceptually  incorrect. The   impact of this  conceptual error depends on the degree to which polarization is sensitive  to collisions. For instance, in regions of the Sun where the  hydrogen density $n_\text{H}$ is low enough to imply that collision effects are negligible, this conceptual error becomes irrelevant. 
  
    Collisions have various effects on the polarized lines, including broadening, depolarization, and the partial redistribution of frequencies. In our investigation, we focused on one specific aspect, which is the depolarization caused by collisions.  By solving the  SEE   for    K I, which is modeled as a multilevel    atom  with  HFS, we show the  role that collisions can play in the modeling of polarization formation. 
Therefore, one must be careful when using collisional rates calculated in the zero-field case to interpret atomic polarization of   HFS  levels in magnetized media. In the cases where level crossings  take place, such as in solar alkali atoms, close coupling treatments of atomic collisions including magnetic fields are necessary in order to properly decipher the information encoded in the polarized radiation.

   Increasing the strength of the external magnetic field not only induces the mixing of states with different total angular momenta but also exerts an influence on collisional rates through modification of the value of the total Hamiltonian of the system  (e.g.,  Volpi \& Bohn  2002,  Krems \& Dalgarno 2004, Bivona et al. 2005).  To   ensure the unicity and reliability of the solution of      the  problems concerned with the  formation of polarization in spectral lines originating from alkaline atoms, such as K I, all major processes must be included in the modeling.  Proper treatment of the collisions in the conditions of the ({incomplete})   PB regime   might lead to novel interpretations 
      and could allow confirmation of the findings of previous works or rectify their shortcomings.

  \begin{acknowledgements}  This research work was funded by Institutional Fund Projects under grant no. (IFPIP:230-130-1443). The authors gratefully acknowledge technical and financial support provided by the Ministry of Education and King Abdulaziz University, DSR, Jeddah, Saudi Arabia. \end{acknowledgements}

\newpage
\begin{appendix}
\section{The statistical Equilibrium equations}
In the system of equations below, $\rho^k_q(L_J,F\,F')$ denote the atomic density matrix elements describing  the hyperfine structure levels $F$ and $F'$ for the level with orbital angular momentum $L$, electronic angular momentum $J$ and nuclear spin $I$. The total angular momentum $F$   takes values between $|J\!-\!I|$ and $J\!+\!I$;  $k$ denotes the tensorial order of the atomic density matrix element. $\dot{\rho}$ represents the variation in $\rho$ with 
time.
$A(L'_{J'} \!\!\rightarrow\!\! L_{J})$ and $B(L_{J} \!\!\rightarrow\!\! L'_{J'})$ denote Einstein's coefficients for spontaneous emission and absorption, respectively. $J^{k_r}_{q_r}(\nu_{L'_{J'},L_{J}})$ indicates the value of the radiation field tensor of order ${k_r}$ at the frequency $\nu_{L'_{J'},L_{J}}$. $D^k(L_{J},F\,F')$ and $D^k(L_{J},F\,F'\!\!\rightarrow\!\! L'_{J'},F\,F')$ respectively denote the collisional relaxation and the collisional transfer  rates affecting the matrix element.  In order to calculate the $\rho^k_q(L_J,F\,F')$ unknowns,  we have
assumed statistical equilibrium, i.e., $\dot{\rho}^k_q(L_J,F\,F')$=0. It is important to note that since the resulting system of equations is
not linearly independent,  one of the equations associated with the population of the sublevels, typically $\dot{\rho }^0_0 (S_{\frac{1}{2}},1,1)=0$, which corresponds to the ground sublevel population, must be replaced by the trace equation $\sum_{i}\sqrt{2F_i+1}\rho_0^0(F_i) = 1$.
\begin{widetext}
\begin{eqnarray}
\dot{\rho }^0_0 (S_{\frac{1}{2}},1\,1)
\!\!\!&=&\!\!\!
-\Big[ 
B(S_{\frac{1}{2}} \!\!\rightarrow\!\! P_{\frac{1}{2}}) J^{0}_{0}(\nu_{{P_{\frac{1}{2}},S_{\frac{1}{2}}}})
+ B(S_{\frac{1}{2}} \!\!\rightarrow\!\! P_{\frac{3}{2}}) J^{0}_{0}(\nu_{{P_{\frac{3}{2}},S_{\frac{1}{2}}}}) 
+ D^0(S_{\frac{1}{2}},1\, 1)
\Big] \rho^0_0(S_{\frac{1}{2}},1\, 1) 
 \\  &&\!\!\!
+\frac{1}{6} A(P_{\frac{1}{2}} \!\!\rightarrow\!\! S_{\frac{1}{2}}) \rho^0_0(P_{\frac{1}{2}},1\, 1) + D^0(S_{\frac{1}{2}},2\,2 \!\!\rightarrow\!\! S_{\frac{1}{2}},1\, 1) \rho^0_0(S_{\frac{1}{2}},2\, 2)
+\frac{1}{2} \sqrt{\frac{5}{3}} A(P_{\frac{1}{2}} \!\!\rightarrow\!\! S_{\frac{1}{2}}) \rho^0_0(P_{\frac{1}{2}},2\, 2)
\nonumber \\  &&\!\!\!
+\frac{1}{\sqrt{3}} A(P_{\frac{3}{2}} \!\!\rightarrow\!\! S_{\frac{1}{2}}) \rho^0_0(P_{\frac{3}{2}},0\, 0) +\frac{5}{6} A(P_{\frac{3}{2}} \!\!\rightarrow\!\! S_{\frac{1}{2}}) \rho^0_0(P_{\frac{3}{2}},1\, 1)
+\frac{1}{2} \sqrt{\frac{5}{3}} A(P_{\frac{3}{2}} \!\!\rightarrow\!\! S_{\frac{1}{2}}) \rho^0_0(P_{\frac{3}{2}},2\, 2) \nonumber
\end{eqnarray}
\begin{eqnarray}
\dot{\rho}^2_0(S_{\frac{1}{2}},1\, 1)
\!\!\!&=&\!\!\!
-\Big[
  B(S_{\frac{1}{2}} \!\!\rightarrow\!\! P_{\frac{1}{2}}) J^{0}_{0}(\nu_{P_{\frac{1}{2}},S_{\frac{1}{2}}})  
+ B(S_{\frac{1}{2}} \!\!\rightarrow\!\! P_{\frac{3}{2}}) J^{0}_{0}(\nu_{P_{\frac{3}{2}},S_{\frac{1}{2}}}) 
+ D^2(S_{\frac{1}{2}},1\,1) 
\Big] \rho^2_0(S_{\frac{1}{2}},1\, 1)
  \\  &&\!\!\!
-\frac{1}{12} A(P_{\frac{1}{2}} \!\!\rightarrow\!\! S_{\frac{1}{2}}) \rho^2_0(P_{\frac{1}{2}},1\,1)+D^2(S_{\frac{1}{2}},2\,2 \!\!\rightarrow\!\! S_{\frac{1}{2}},1\,1) \rho^2_0(S_{\frac{1}{2}},2\,2)
+\frac{1}{4} A(P_{\frac{1}{2}} \!\!\rightarrow\!\! S_{\frac{1}{2}}) \rho^2_0(P_{\frac{1}{2}},1\,2)
\nonumber \\  &&\!\!\!
-\frac{1}{4} A(P_{\frac{1}{2}} \!\!\rightarrow\!\! S_{\frac{1}{2}}) \rho^2_0(P_{\frac{1}{2}},2\,1)+\frac{1}{4}\sqrt{\frac{7}{3}} A(P_{\frac{1}{2}} \!\!\rightarrow\!\! S_{\frac{1}{2}}) \rho^2_0(P_{\frac{1}{2}},2\,2)
+\frac{1}{\sqrt{6}} A(P_{\frac{3}{2}} \!\!\rightarrow\!\! S_{\frac{1}{2}}) \rho^2_0(P_{\frac{3}{2}},0\,2)
\nonumber \\  &&\!\!\!
-\frac{5}{12} A(P_{\frac{3}{2}} \!\!\rightarrow\!\! S_{\frac{1}{2}}) \rho^2_0(P_{\frac{3}{2}},1\,1)+\frac{\sqrt{5}}{4}  A(P_{\frac{3}{2}} \!\!\rightarrow\!\! S_{\frac{1}{2}}) \rho^2_0(P_{\frac{3}{2}},1\,2)
+\frac{1}{\sqrt{6}} A(P_{\frac{3}{2}} \!\!\rightarrow\!\! S_{\frac{1}{2}}) \rho^2_0(P_{\frac{3}{2}},2\,0)
\nonumber \\  &&\!\!\!
-\frac{\sqrt{5}}{4}  A(P_{\frac{3}{2}} \!\!\rightarrow\!\! S_{\frac{1}{2}}) \rho^2_0(P_{\frac{3}{2}},2\,1)+\frac{1}{4} \sqrt{\frac{7}{3}} A(P_{\frac{3}{2}} \!\!\rightarrow\!\! S_{\frac{1}{2}}) \rho^2_0(P_{\frac{3}{2}},2\,2) \nonumber
\end{eqnarray}
\begin{eqnarray}
\dot{\rho}^2_0(S_{\frac{1}{2}},1\,2)
\!\!\!&=&\!\!\!
-\Big[
  B(S_{\frac{1}{2}} \!\!\rightarrow\!\! P_{\frac{1}{2}})  J^{0}_{0}(\nu_{P_{\frac{1}{2}},S_{\frac{1}{2}}})  
+ B(S_{\frac{1}{2}} \!\!\rightarrow\!\! P_{\frac{3}{2}})  J^{0}_{0}(\nu_{P_{\frac{3}{2}},S_{\frac{1}{2}}})  
+ D^2(S_{\frac{1}{2}},1\,2) 
\Big] \rho^2_0(S_{\frac{1}{2}},1\,2)
 \\  &&\!\!\!\!
+\frac{1}{4} A(P_{\frac{1}{2}} \!\!\rightarrow\!\! S_{\frac{1}{2}}) \rho^2_0(P_{\frac{1}{2}},1\,1) -\frac{1}{12} A(P_{\frac{1}{2}} \!\!\rightarrow\!\! S_{\frac{1}{2}}) \rho^2_0(P_{\frac{1}{2}},1\,2)
-\frac{1}{4}  A(P_{\frac{1}{2}} \!\!\rightarrow\!\! S_{\frac{1}{2}}) \rho^2_0(P_{\frac{1}{2}},2\,1)
\nonumber \\  &&\!\!\!\!
+\frac{1}{4}  \sqrt{\frac{7}{3}} A(P_{\frac{1}{2}} \!\!\rightarrow\!\! S_{\frac{1}{2}}) \rho^2_0(P_{\frac{1}{2}},2\,2)+\frac{1}{\sqrt{6}} A(P_{\frac{3}{2}} \!\!\rightarrow\!\! S_{\frac{1}{2}}) \rho^2_0(P_{\frac{3}{2}},0\,2)
-\frac{1}{4}  A(P_{\frac{3}{2}} \!\!\rightarrow\!\! S_{\frac{1}{2}}) \rho^2_0(P_{\frac{3}{2}},1\,1)
\nonumber \\  &&\!\!\!\!
+\frac{\sqrt{5}}{12}  A(P_{\frac{3}{2}}  \!\!\rightarrow\!\!  S_{\frac{1}{2}}) \rho^2_0(P_{\frac{3}{2}},1\,2)+\frac{1}{3} \sqrt{\frac{7}{2}} A(P_{\frac{3}{2}} \!\!\rightarrow\!\! S_{\frac{1}{2}}) \rho^2_0(P_{\frac{3}{2}},1\,3)
+\frac{1}{4 \sqrt{5}} A(P_{\frac{3}{2}} \!\!\rightarrow\!\! S_{\frac{1}{2}}) \rho^2_0(P_{\frac{3}{2}},2\,1)
\nonumber \\  &&\!\!\!\!
-\frac{1}{4} \sqrt{\frac{7}{3}} A(P_{\frac{3}{2}} \!\!\rightarrow\!\! S_{\frac{1}{2}}) \rho^2_0(P_{\frac{3}{2}},2\,2)+\sqrt{\frac{7}{15}} A(P_{\frac{3}{2}} \!\!\rightarrow\!\! S_{\frac{1}{2}}) \rho^2_0(P_{\frac{3}{2}},2\,3) \nonumber
\end{eqnarray}
\begin{eqnarray}
\dot{\rho}^2_0(S_{\frac{1}{2}},2\,1)
\!\!\!&=&\!\!\!
- \Big[
  B(S_{\frac{1}{2}} \!\!\rightarrow\!\! P_{\frac{1}{2}}) J_0^0(\nu_{P_{\frac{1}{2}},S_{\frac{1}{2}}})
+ B(S_{\frac{1}{2}} \!\!\rightarrow\!\! P_{\frac{3}{2}})  J_0^0(\nu_{P_{\frac{3}{2}},S_{\frac{1}{2}}})
+ D^2(S_{\frac{1}{2}},2\,1) 
\Big] \rho^2_0(S_{\frac{1}{2}},2\,1)
 \\  &&\!\!\!
- \frac{1}{4} A(P_{\frac{1}{2}} \!\!\rightarrow\!\! S_{\frac{1}{2}}) \rho^2_0(P_{\frac{1}{2}},1\,1)- \frac{1}{4} A(P_{\frac{1}{2}} \!\!\rightarrow\!\! S_{\frac{1}{2}}) \rho^2_0(P_{\frac{1}{2}},1\,2)
- \frac{1}{12} A(P_{\frac{1}{2}} \!\!\rightarrow\!\! S_{\frac{1}{2}}) \rho^2_0(P_{\frac{1}{2}},2\,1)
\nonumber \\  &&\!\!\!
- \frac{1}{4} \sqrt{\frac{7}{3}} A(P_{\frac{1}{2}} \!\!\rightarrow\!\! S_{\frac{1}{2}}) \rho^2_0(P_{\frac{1}{2}},2\,2)+ \frac{1}{4} A(P_{\frac{3}{2}} \!\!\rightarrow\!\! S_{\frac{1}{2}}) \rho^2_0(P_{\frac{3}{2}},1\,1)
+ \frac{1}{4 \sqrt{5}} A(P_{\frac{3}{2}} \!\!\rightarrow\!\! S_{\frac{1}{2}}) \rho^2_0(P_{\frac{3}{2}},1\,2)
\nonumber \\  &&\!\!\!
- \frac{1}{\sqrt{6}} A(P_{\frac{3}{2}} \!\!\rightarrow\!\! S_{\frac{1}{2}}) \rho^2_0(P_{\frac{3}{2}},2\,0)+ \frac{\sqrt{5}}{12}  A(P_{\frac{3}{2}} \!\!\rightarrow\!\! S_{\frac{1}{2}}) \rho^2_0(P_{\frac{3}{2}},2\,1)
+ \frac{1}{4} \sqrt{\frac{7}{3}} A(P_{\frac{3}{2}} \!\!\rightarrow\!\! S_{\frac{1}{2}}) \rho^2_0(P_{\frac{3}{2}},2\,2)
\nonumber \\  &&\!\!\!
- \frac{1}{3} \sqrt{\frac{7}{2}} A(P_{\frac{3}{2}} \!\!\rightarrow\!\! S_{\frac{1}{2}}) \rho^2_0(P_{\frac{3}{2}},3\,1)+ \sqrt{\frac{7}{15}} A(P_{\frac{3}{2}} \!\!\rightarrow\!\! S_{\frac{1}{2}}) \rho^2_0(P_{\frac{3}{2}},3\,2) \nonumber
\end{eqnarray}
\begin{eqnarray}
\dot{\rho}^0_0(S_{\frac{1}{2}},2\,2)
\!\!\!&=&\!\!\!
- \Big[
  B(S_{\frac{1}{2}} \!\!\rightarrow\!\! P_{\frac{1}{2}}) J_0^0(\nu_{P_{\frac{1}{2}},S_{\frac{1}{2}}})  
+ B(S_{\frac{1}{2}} \!\!\rightarrow\!\! P_{\frac{3}{2}}) J_0^0(\nu_{P_{\frac{3}{2}},S_{\frac{1}{2}}}) 
+ D^0(S_{\frac{1}{2}},2\,2) 
\Big]    \rho^0_0(S_{\frac{1}{2}},2\,2)
 \\  &&\!\!\!
+ \frac{1}{2} \sqrt{\frac{5}{3}} A(P_{\frac{1}{2}} \!\!\rightarrow\!\! S_{\frac{1}{2}}) \rho^0_0(P_{\frac{1}{2}},1\,1)+ \frac{1}{2} A(P_{\frac{1}{2}} \!\!\rightarrow\!\! S_{\frac{1}{2}}) \rho^0_0(P_{\frac{1}{2}},2\,2)
+ \frac{1}{2 \sqrt{15}} A(P_{\frac{3}{2}} \!\!\rightarrow\!\! S_{\frac{1}{2}}) \rho^0_0(P_{\frac{3}{2}},1\,1)
\nonumber \\  &&\!\!\!
+ \frac{1}{2} A(P_{\frac{3}{2}} \!\!\rightarrow\!\! S_{\frac{1}{2}}) \rho^0_0(P_{\frac{3}{2}},2\,2)+ \sqrt{\frac{7}{5}} A(P_{\frac{3}{2}} \!\!\rightarrow\!\! S_{\frac{1}{2}}) \rho^0_0(P_{\frac{3}{2}},3\,3)
+ D^0(S_{\frac{1}{2}},1\,1 \!\!\rightarrow\!\! S_{\frac{1}{2}},2\,2) \rho^0_0(S_{\frac{1}{2}},1\,1) \nonumber
\end{eqnarray}
\begin{eqnarray}
\dot{\rho}^2_0(S_{\frac{1}{2}},2\,2)
\!\!\!&=&\!\!\!
- \Big[
  B(S_{\frac{1}{2}} \!\!\rightarrow\!\! P_{\frac{1}{2}}) J_0^0(\nu_{P_{\frac{1}{2}},S_{\frac{1}{2}}}) 
+ B(S_{\frac{1}{2}} \!\!\rightarrow\!\! P_{\frac{3}{2}}) J_0^0(\nu_{P_{\frac{3}{2}},S_{\frac{1}{2}}}) 
+ D^2(S_{\frac{1}{2}},2\,2) 
\Big]  \rho^2_0(S_{\frac{1}{2}},2\,2) 
  \\  &&\!\!\!
+ \frac{1}{4} \sqrt{\frac{7}{3}} A(P_{\frac{1}{2}} \!\!\rightarrow\!\! S_{\frac{1}{2}}) \rho^2_0(P_{\frac{1}{2}},1\,1)+ \frac{1}{4} \sqrt{\frac{7}{3}} A(P_{\frac{1}{2}} \!\!\rightarrow\!\! S_{\frac{1}{2}}) \rho^2_0(P_{\frac{1}{2}},1\,2)
- \frac{1}{4} \sqrt{\frac{7}{3}} A(P_{\frac{1}{2}} \!\!\rightarrow\!\! S_{\frac{1}{2}}) \rho^2_0(P_{\frac{1}{2}},2\,1)
\nonumber \\  &&\!\!\!
+ \frac{1}{4} A(P_{\frac{1}{2}} \!\!\rightarrow\!\! S_{\frac{1}{2}}) \rho^2_0(P_{\frac{1}{2}},2\,2)+ \frac{1}{20} \sqrt{\frac{7}{3}} A(P_{\frac{3}{2}} \!\!\rightarrow\!\! S_{\frac{1}{2}}) \rho^2_0(P_{\frac{3}{2}},1\,1)
+ \frac{1}{4} \sqrt{\frac{7}{15}} A(P_{\frac{3}{2}} \!\!\rightarrow\!\! S_{\frac{1}{2}}) \rho^2_0(P_{\frac{3}{2}},1\,2)
\nonumber \\  &&\!\!\!
+ \frac{1}{5 \sqrt{6}} A(P_{\frac{3}{2}} \!\!\rightarrow\!\! S_{\frac{1}{2}}) \rho^2_0(P_{\frac{3}{2}},1\,3)- \frac{1}{4} \sqrt{\frac{7}{15}} A(P_{\frac{3}{2}} \!\!\rightarrow\!\! S_{\frac{1}{2}}) \rho^2_0(P_{\frac{3}{2}},2\,1)
+ \frac{1}{4} A(P_{\frac{3}{2}}  \!\!\rightarrow\!\!  S_{\frac{1}{2}}) \rho^2_0(P_{\frac{3}{2}},2\,2)
\nonumber \\  &&\!\!\!
+ \frac{1}{\sqrt{5}} A(P_{\frac{3}{2}} \!\!\rightarrow\!\! S_{\frac{1}{2}}) \rho^2_0(P_{\frac{3}{2}},2\,3)+ \frac{1}{5\sqrt{6}} A(P_{\frac{3}{2}} \!\!\rightarrow\!\! S_{\frac{1}{2}}) \rho^2_0(P_{\frac{3}{2}},3\,1)
- \frac{1}{\sqrt{5}} A(P_{\frac{3}{2}} \!\!\rightarrow\!\! S_{\frac{1}{2}}) \rho^2_0(P_{\frac{3}{2}},3\,2)
\nonumber \\  &&\!\!\!
+\frac{2\sqrt{6}}{5}  A(P_{\frac{3}{2}} \!\!\rightarrow\!\! S_{\frac{1}{2}}) \rho^2_0(P_{\frac{3}{2}},3\,3)+ D^2(S_{\frac{1}{2}},1\,1 \!\!\rightarrow\!\! S_{\frac{1}{2}},2\,2) \rho^2_0(S_{\frac{1}{2}},1\,1) \nonumber
\end{eqnarray}
\begin{eqnarray}
\dot{\rho}^4_0(S_{\frac{1}{2}},2\,2)
\!\!\!&=&\!\!\!
- \Big[
  B(S_{\frac{1}{2}} \!\!\rightarrow\!\! P_{\frac{1}{2}})  J_0^0(\nu_{P_{\frac{1}{2}},S_{\frac{1}{2}}})
+ B(S_{\frac{1}{2}} \!\!\rightarrow\!\! P_{\frac{3}{2}})  J_0^0(\nu_{P_{\frac{3}{2}},S_{\frac{1}{2}}})
+ D^4(S_{\frac{1}{2}},2\,2) 
\Big]  \rho^4_0(S_{\frac{1}{2}},2\,2) 
 \\  &&\!\!\!
- \frac{1}{3} A(P_{\frac{1}{2}} \!\!\rightarrow\!\! S_{\frac{1}{2}}) \rho^4_0(P_{\frac{1}{2}},2\,2)+ \frac{1}{\sqrt{10}} A(P_{\frac{3}{2}} \!\!\rightarrow\!\! S_{\frac{1}{2}}) \rho^4_0(P_{\frac{3}{2}},1\,3)
- \frac{1}{3} A(P_{\frac{3}{2}} \!\!\rightarrow\!\! S_{\frac{1}{2}}) \rho^4_0(P_{\frac{3}{2}},2\,2)
\nonumber \\  &&\!\!\!
+ \frac{1}{3} \sqrt{\frac{5}{2}} A(P_{\frac{3}{2}} \!\!\rightarrow\!\! S_{\frac{1}{2}}) \rho^4_0(P_{\frac{3}{2}},2\,3)+ \frac{1}{\sqrt{10}} A(P_{\frac{3}{2}} \!\!\rightarrow\!\! S_{\frac{1}{2}}) \rho^4_0(P_{\frac{3}{2}},3\,1)
- \frac{1}{3} \sqrt{\frac{5}{2}} A(P_{\frac{3}{2}}  \!\!\rightarrow\!\!  S_{\frac{1}{2}}) \rho^4_0(P_{\frac{3}{2}},3\,2) \nonumber \\  &&
+ \frac{1}{3} \sqrt{\frac{11}{5}} A(P_{\frac{3}{2}} \!\!\rightarrow\!\! S_{\frac{1}{2}}) \rho^4_0(P_{\frac{3}{2}},3\,3)  \nonumber
\end{eqnarray}
\begin{eqnarray}
\dot{\rho}^0_0(P_{\frac{1}{2}},1\,1)
\!\!\!&=&\!\!\!
- \Big[
  A(P_{\frac{1}{2}} \!\!\rightarrow\!\! S_{\frac{1}{2}}) 
+ D^0(P_{\frac{1}{2}},1\,1) 
+ D^0(P_{\frac{1}{2}},1\,1 \!\!\rightarrow\!\! P_{\frac{3}{2}},1\,1) 
  \Big]   \rho^0_0(P_{\frac{1}{2}},1\,1)
 \\  &&\!\!\!
+ B(S_{\frac{1}{2}} \!\!\rightarrow\!\! P_{\frac{1}{2}})
\Bigg[
J_0^0(\nu_{P_{\frac{1}{2}},S_{\frac{1}{2}}})
\left\{
  \frac{1}{6}   \rho^0_0(S_{\frac{1}{2}},1\,1) 
+ \frac{1}{2} \sqrt{\frac{5}{3}}  \rho^0_0(S_{\frac{1}{2}},2\,2) 
\right\}
\nonumber \\  &&\!\!\!
+ J^2_0(\nu_{P_{\frac{1}{2}},S_{\frac{1}{2}}})
\left\{
- \frac{1}{12}   \rho^2_0(S_{\frac{1}{2}},1\,1) 
- \frac{1}{4}     \rho^2_0(S_{\frac{1}{2}},1\,2)
+ \frac{1}{4}    \rho^2_0(S_{\frac{1}{2}},2\,1) 
+ \frac{1}{4} \sqrt{\frac{7}{3}}   \rho^2_0(S_{\frac{1}{2}},2\,2) 
\right\}
\Bigg]  
\nonumber \\  &&\!\!\!
+ D^0(P_{\frac{1}{2}},2\,2 \!\!\rightarrow\!\! P_{\frac{1}{2}},1\,1) \rho^0_0(P_{\frac{1}{2}},2\,2)
+ D^0(P_{\frac{3}{2}},0\,0 \!\!\rightarrow\!\! P_{\frac{1}{2}},1\,1) \rho^0_0(P_{\frac{3}{2}},0\,0)
+ D^0(P_{\frac{3}{2}},1\,1 \!\!\rightarrow\!\! P_{\frac{1}{2}},1\,1) \rho^0_0(P_{\frac{3}{2}},1\,1)
\nonumber \\  &&\!\!\!
+ D^0(P_{\frac{3}{2}},2\,2 \!\!\rightarrow\!\! P_{\frac{1}{2}},1\,1) \rho^0_0(P_{\frac{3}{2}},2\,2) 
+ D^0(P_{\frac{3}{2}},3\,3 \!\!\rightarrow\!\! P_{\frac{1}{2}},1\,1) \rho^0_0(P_{\frac{3}{2}},3\,3) \nonumber
\end{eqnarray}
\begin{eqnarray}
\dot{\rho}^2_0(P_{\frac{1}{2}},1\,1)
\!\!\!&=&\!\!\!
- \Big[
  A(P_{\frac{1}{2}} \!\!\rightarrow\!\! S_{\frac{1}{2}}) 
+ D^2(P_{\frac{1}{2}},1\,1) 
+ D^0(P_{\frac{1}{2}},1\,1 \!\!\rightarrow\!\! P_{\frac{3}{2}},1\,1)
  \Big] \rho^2_0(P_{\frac{1}{2}},1\,1)
+ B(S_{\frac{1}{2}} \!\!\rightarrow\!\! P_{\frac{1}{2}})
\Bigg[ 
J_0^0(\nu_{P_{\frac{1}{2}},S_{\frac{1}{2}}})
 \\  &&\!\!\!
\times
\!\left\{\!
- \frac{1}{12}  \rho^2_0(S_{\frac{1}{2}},1\,1)
+ \frac{1}{4}  \rho^2_0(S_{\frac{1}{2}},1\,2) 
- \frac{1}{4}  \rho^2_0(S_{\frac{1}{2}},2\,1) 
+ \frac{1}{4} \sqrt{\frac{7}{3}}  \rho^2_0(S_{\frac{1}{2}},2\,2) 
- \frac{1}{4}  \rho^2_0(S_{\frac{1}{2}},2\,1) 
+ \frac{1}{4} \sqrt{\frac{7}{3}}  \rho^2_0(S_{\frac{1}{2}},2\,2) 
\!\right\}
\nonumber \\  &&\!\!\!
+ J^2_0(\nu_{P_{\frac{1}{2}},S_{\frac{1}{2}}})
\!\left\{\!
\!-\! \frac{1}{12}  \rho^0_0(S_{\frac{1}{2}},1\,1) 
\!-\! \frac{1}{6 \sqrt{2}} \rho^2_0(S_{\frac{1}{2}},1\,1) 
\!+\! \frac{1}{4 \sqrt{15}} \rho^0_0(S_{\frac{1}{2}},2\,2)
\!+\! \frac{1}{2 \sqrt{42}} \rho^2_0(S_{\frac{1}{2}},2\,2) 
\!+\! 3 \sqrt{\frac{3}{70}}  \rho^4_0(S_{\frac{1}{2}},2\,2) 
\!\right\}\!
\!\Bigg]\!
\nonumber \\  &&\!\!\!
+ D^2(P_{\frac{1}{2}},2\,2 \!\!\rightarrow\!\! P_{\frac{1}{2}},1\,1) \rho^2_0(P_{\frac{1}{2}},2\,2)
+ D^2(P_{\frac{3}{2}},1\,1 \!\!\rightarrow\!\! P_{\frac{1}{2}},1\,1) \rho^2_0(P_{\frac{3}{2}},1\,1)
+ D^2(P_{\frac{3}{2}},2\,2 \!\!\rightarrow\!\! P_{\frac{1}{2}},1\,1) \rho^2_0(P_{\frac{3}{2}},2\,2)
\nonumber \\  &&\!\!\!
+ D^2(P_{\frac{3}{2}},3\,3 \!\!\rightarrow\!\! P_{\frac{1}{2}},1\,1) \rho^2_0(P_{\frac{3}{2}},3\,3) \nonumber
\end{eqnarray}
\begin{eqnarray}
\dot{\rho}^2_0(P_{\frac{1}{2}},1\,2)
\!\!\!&=&\!\!\!
- \Big[
  A(P_{\frac{1}{2}} \!\!\rightarrow\!\! S_{\frac{1}{2}}) 
+ D^2(P_{\frac{1}{2}},1\,2) 
+ D^0(P_{\frac{1}{2}},1\,2 \!\!\rightarrow\!\! P_{\frac{3}{2}},1\,2) 
\Big]  \rho^2_0(P_{\frac{1}{2}},1\,2) 
+ B(S_{\frac{1}{2}} \!\!\rightarrow\!\! P_{\frac{1}{2}})
\times
\\  &&\!\!\!
\Bigg[
J_0^0(\nu_{P_{\frac{1}{2}},S_{\frac{1}{2}}})
\left\{
  \frac{1}{4}  \rho^2_0(S_{\frac{1}{2}},1\,1) 
- \frac{1}{12}  \rho^2_0(S_{\frac{1}{2}},1\,2) 
- \frac{1}{4}  \rho^2_0(S_{\frac{1}{2}},2\,1) 
+ \frac{1}{4} \sqrt{\frac{7}{3}}  \rho^2_0(S_{\frac{1}{2}},2\,2) 
\right\}
\nonumber \\  &&\!\!\!
+ J^2_0(\nu_{P_{\frac{1}{2}},S_{\frac{1}{2}}})
\left\{
- \frac{1}{4}  \rho^0_0(S_{\frac{1}{2}},1\,1)
+ \frac{1}{4} \sqrt{\frac{3}{5}}  \rho^0_0(S_{\frac{1}{2}},2\,2) 
+ \frac{1}{6 \sqrt{2}} \rho^2_0(S_{\frac{1}{2}},1\,2)
\right.
\nonumber \\  &&\!\!\! 
\left.
+ \frac{1}{2 \sqrt{2}} \rho^2_0(S_{\frac{1}{2}},2\,1) 
+ \frac{1}{\sqrt{42}} \rho^2_0(S_{\frac{1}{2}},2\,2) 
- \sqrt{\frac{3}{70}}  \rho^4_0(S_{\frac{1}{2}},2\,2) 
\right\}
\Bigg] \nonumber
\end{eqnarray}
\begin{eqnarray}
\dot{\rho}^2_0(P_{\frac{1}{2}},2\,1)
\!\!\!&=&\!\!\!
- \Big[
  A(P_{\frac{1}{2}} \!\!\rightarrow\!\! S_{\frac{1}{2}}) 
+ D^2(P_{\frac{1}{2}},2\,1)
+ D^0(P_{\frac{1}{2}},2\,1 \!\!\rightarrow\!\! P_{\frac{3}{2}},2\,1) 
\Big]  \rho^2_0(P_{\frac{1}{2}},2\,1) 
 \\  &&\!\!\!
+ B(S_{\frac{1}{2}} \!\!\rightarrow\!\! P_{\frac{1}{2}})
\Bigg[
- J_0^0(\nu_{P_{\frac{1}{2}},S_{\frac{1}{2}}})
\left\{
  \frac{1}{4} \rho^2_0(S_{\frac{1}{2}},1\,1) 
+ \frac{1}{4}  \rho^2_0(S_{\frac{1}{2}},1\,2) 
+ \frac{1}{12}  \rho^2_0(S_{\frac{1}{2}},2\,1) 
+ \frac{1}{4} \sqrt{\frac{7}{3}}  \rho^2_0(S_{\frac{1}{2}},2\,2) 
\right\}
\nonumber \\  &&\!\!\! 
+ J^2_0(\nu_{P_{\frac{1}{2}},S_{\frac{1}{2}}})
\left\{
  \frac{1}{4}  \rho^0_0(S_{\frac{1}{2}},1\,1) 
- \frac{1}{4} \sqrt{\frac{3}{5}}  \rho^0_0(S_{\frac{1}{2}},2\,2) 
+ \frac{1}{2 \sqrt{2}} \rho^2_0(S_{\frac{1}{2}},1\,2) 
\right. 
\nonumber \\  &&\!\!\! 
\left.
+ \frac{1}{6 \sqrt{2}} \rho^2_0(S_{\frac{1}{2}},2\,1) 
- \frac{1}{\sqrt{42}} \rho^2_0(S_{\frac{1}{2}},2\,2) 
+ \sqrt{\frac{3}{70}}  \rho^4_0(S_{\frac{1}{2}},2\,2) 
\right\}
\Bigg] \nonumber
\end{eqnarray}
\begin{eqnarray}
\dot{\rho}^0_0(P_{\frac{1}{2}},2\,2)
\!\!\!&=&\!\!\!
- \Big[
  A(P_{\frac{1}{2}} \!\!\rightarrow\!\! S_{\frac{1}{2}}) 
+ D^0(P_{\frac{1}{2}},2\,2)
+ D^0(P_{\frac{1}{2}},2\,2 \!\!\rightarrow\!\! P_{\frac{3}{2}},2\,2) 
\Big]  \rho^0_0(P_{\frac{1}{2}},2\,2) 
 \\  &&\!\!\!
+ B(S_{\frac{1}{2}} \!\!\rightarrow\!\! P_{\frac{1}{2}})
\Bigg[
J_0^0(\nu_{P_{\frac{1}{2}},S_{\frac{1}{2}}})
\left\{
  \frac{1}{2} \sqrt{\frac{5}{3}}  \rho^0(S_{\frac{1}{2}},1\,1) 
+ \frac{1}{2}  \rho^0(S_{\frac{1}{2}},2\,2) 
\right\} 
\nonumber \\  &&\!\!\! 
+ J^2_0(\nu_{P_{\frac{1}{2}},S_{\frac{1}{2}}})
\left\{
  \frac{1}{4 \sqrt{15}} \rho^2_0(S_{\frac{1}{2}},1\,1) 
+ \frac{1}{4} \sqrt{\frac{3}{5}}  \rho^2_0(S_{\frac{1}{2}},1\,2) 
- \frac{1}{4} \sqrt{\frac{3}{5}}  \rho^2_0(S_{\frac{1}{2}},2\,1) 
- \frac{1}{4} \sqrt{\frac{7}{5}}  \rho^2_0(S_{\frac{1}{2}},2\,2) 
\right\}
\Bigg]
\nonumber \\  &&\!\!\!
+ D^0(P_{\frac{1}{2}},1\,1 \!\!\rightarrow\!\! P_{\frac{1}{2}},2\,2) \rho^0_0(P_{\frac{1}{2}},1\,1)
+ D^0(P_{\frac{3}{2}},0\,0 \!\!\rightarrow\!\! P_{\frac{1}{2}},2\,2) \rho^0_0(P_{\frac{3}{2}},0\,0)
+ D^0(P_{\frac{3}{2}},1\,1 \!\!\rightarrow\!\! P_{\frac{1}{2}},2\,2) \rho^0_0(P_{\frac{3}{2}},1\,1)
\nonumber \\  &&\!\!\!
+ D^0(P_{\frac{3}{2}},2\,2 \!\!\rightarrow\!\! P_{\frac{1}{2}},2\,2) \rho^0_0(P_{\frac{3}{2}},2\,2)
+ D^0(P_{\frac{3}{2}},3\,3 \!\!\rightarrow\!\! P_{\frac{1}{2}},2\,2) \rho^0_0(P_{\frac{3}{2}},3\,3) \nonumber 
\end{eqnarray}
\begin{eqnarray}
\dot{\rho}^2_0(P_{\frac{1}{2}},2\,2)
\!\!\!&=&\!\!\!
- \Big[
   A(P_{\frac{1}{2}} \!\!\rightarrow\!\! S_{\frac{1}{2}}) 
+  D^2(P_{\frac{1}{2}},2\,2)
+  D^0(P_{\frac{1}{2}},2\,2 \!\!\rightarrow\!\! P_{\frac{3}{2}},2\,2) 
\Big]   \rho^2_0(P_{\frac{1}{2}},2\,2)  
 \\  &&\!\!\!
+ B(S_{\frac{1}{2}} \!\!\rightarrow\!\! P_{\frac{1}{2}}) 
\Bigg[
J_0^0(\nu_{P_{\frac{1}{2}},S_{\frac{1}{2}}})
\left\{
    \frac{1}{4} \sqrt{\frac{7}{3}} \rho_0^2 (S_{\frac{1}{2}},1\,1) 
+   \frac{1}{4} \sqrt{\frac{7}{3}}  \rho_2^2 (S_{\frac{1}{2}},1\,2)
-  \frac{1}{4} \sqrt{\frac{7}{3}}   \rho^2_0(S_{\frac{1}{2}},2\,1)
+   \frac{1}{4}  \rho^2_0(S_{\frac{1}{2}},2\,2)  
\right\}
\nonumber \\  &&\!\!\!
+ J^2_0(\nu_{P_{\frac{1}{2}},S_{\frac{1}{2}}})
\left\{
   \frac{1}{4} \sqrt{\frac{7}{3}} \rho^0_0(S_{\frac{1}{2}},1\,1) 
-  \frac{1}{4} \sqrt{\frac{7}{5}} \rho^0_0(S_{\frac{1}{2}},2\,2)
+  \frac{1}{2 \sqrt{42}} \rho^2_0(S_{\frac{1}{2}},1\,1) 
+  \frac{1}{\sqrt{42}} \rho^2_0(S_{\frac{1}{2}},1\,2) 
\right.
\nonumber \\  &&\!\!\!
\left. 
-  \frac{1}{\sqrt{42}} \rho^2_0(S_{\frac{1}{2}},2\,1)
-  \frac{5}{14 \sqrt{2}}  \rho^2_0(S_{\frac{1}{2}},2\,2) 
-  \frac{3}{7\sqrt{10}}  \rho^4_0(S_{\frac{1}{2}},2\,2) 
\right\}
\Bigg]
\nonumber \\  &&\!\!\!
+  D^2(P_{\frac{1}{2}},1\,1 \!\!\rightarrow\!\! P_{\frac{1}{2}},2\,2) \rho^2_0(P_{\frac{1}{2}},1\,1)
+  D^2(P_{\frac{3}{2}},1\,1 \!\!\rightarrow\!\! P_{\frac{1}{2}},2\,2) \rho^2_0(P_{\frac{3}{2}},1\,1)
+  D^2(P_{\frac{3}{2}},2\,2 \!\!\rightarrow\!\! P_{\frac{1}{2}},2\,2) \rho^2_0(P_{\frac{3}{2}},2\,2)
\nonumber \\  &&\!\!\!
+  D^2(P_{\frac{3}{2}},3\,3 \!\!\rightarrow\!\! P_{\frac{1}{2}},2\,2) \rho^2_0(P_{\frac{3}{2}},3\,3) \nonumber
\end{eqnarray}
\begin{eqnarray}
\dot{\rho}^4_0(P_{\frac{1}{2}},2\,2)
\!\!\!&=&\!\!\!
- \Big[
   A(P_{\frac{1}{2}} \!\!\rightarrow\!\! S_{\frac{1}{2}}) 
+  D^4(P_{\frac{1}{2}},2\,2)
+  D^0(P_{\frac{1}{2}},2\,2 \!\!\rightarrow\!\! P_{\frac{3}{2}},2\,2) 
  \Big]  \rho^4_0(P_{\frac{1}{2}},2\,2) 
  \\  &&\!\!\!
+  B(S_{\frac{1}{2}} \!\!\rightarrow\!\! P_{\frac{1}{2}})
\Bigg[
-   \frac{1}{3}  J_0^0(\nu_{P_{\frac{1}{2}},S_{\frac{1}{2}}})  \rho^4_0(S_{\frac{1}{2}},2\,2) 
+ J^2_0(\nu_{P_{\frac{1}{2}},S_{\frac{1}{2}}})
\times
\nonumber \\  &&\!\!\!
\left\{
 3 \sqrt{\frac{3}{70}}  \rho^2_0(S_{\frac{1}{2}},1\,1) 
-   \sqrt{\frac{3}{70}}  \rho^2_0(S_{\frac{1}{2}},1\,2) 
+   \sqrt{\frac{3}{70}}  \rho^2_0(S_{\frac{1}{2}},2\,1) 
-   \frac{3}{7 \sqrt{10}} \rho^2_0(S_{\frac{1}{2}},2\,2) 
-   \frac{5\sqrt{2}}{21}  \rho^4_0(S_{\frac{1}{2}},2\,2) 
\right\}
\Bigg]
\nonumber \\  &&\!\!\!
+  D^4(P_{\frac{3}{2}},2\,2 \!\!\rightarrow\!\! P_{\frac{1}{2}},2\,2) \rho^4_0(P_{\frac{3}{2}},2\,2)
+  D^4(P_{\frac{3}{2}},3\,3 \!\!\rightarrow\!\! P_{\frac{1}{2}},2\,2) \rho^4_0(P_{\frac{3}{2}},3\,3)  \nonumber
\end{eqnarray}
\begin{eqnarray}
\dot{\rho}^0_0(P_{\frac{3}{2}},0\,0)
\!\!\!&=&\!\!\!
- \Big[
   A(P_{\frac{3}{2}} \!\!\rightarrow\!\! S_{\frac{1}{2}}) 
+  D^0(P_{\frac{3}{2}},0\,0) 
  \Big] \rho^0_0(P_{\frac{3}{2}},0\,0)  
+ B(S_{\frac{1}{2}} \!\!\rightarrow\!\! P_{\frac{3}{2}})
\Big[
 \frac{1}{2\sqrt{3}} J_0^0(\nu_{P_{\frac{3}{2}},S_{\frac{1}{2}}}) \rho^0_0(S_{\frac{1}{2}},1\,1) 
   \\  &&\!\!\!
+\frac{1}{2\sqrt{3}} J^2_0(\nu_{P_{\frac{3}{2}},S_{\frac{1}{2}}}) \rho^2_0(S_{\frac{1}{2}},1\,1) 
\Big]
+D^0(P_{\frac{1}{2}},1\,1 \!\!\rightarrow\!\! P_{\frac{3}{2}},0\,0) \rho^0_0(P_{\frac{1}{2}},1\,1)
+D^0(P_{\frac{1}{2}},2\,2 \!\!\rightarrow\!\! P_{\frac{3}{2}},0\,0) \rho^0_0(P_{\frac{1}{2}},2\,2)
\nonumber \\  &&\!\!\!
+D^0(P_{\frac{3}{2}},1\,1 \!\!\rightarrow\!\! P_{\frac{3}{2}},0\,0) \rho^0_0(P_{\frac{3}{2}},1\,1)
+D^0(P_{\frac{3}{2}},2\,2 \!\!\rightarrow\!\! P_{\frac{3}{2}},0\,0) \rho^0_0(P_{\frac{3}{2}},2\,2)
+D^0(P_{\frac{3}{2}},3\,3 \!\!\rightarrow\!\! P_{\frac{3}{2}},0\,0) \rho^0_0(P_{\frac{3}{2}},3\,3)
\nonumber 
\end{eqnarray}
\begin{eqnarray}
\dot{\rho}^2_0(P_{\frac{3}{2}},0\,2)
\!\!\!&=&\!\!\!
- \Big[
   A(P_{\frac{3}{2}} \!\!\rightarrow\!\! S_{\frac{1}{2}}) 
+  D^2(P_{\frac{3}{2}},0\,2) 
  \Big]  \rho^2_0(P_{\frac{3}{2}},0\,2)  
+  B(S_{\frac{1}{2}} \!\!\rightarrow\!\! P_{\frac{3}{2}})
\Bigg[
J_0^0(\nu_{P_{\frac{3}{2}},S_{\frac{1}{2}}})
\left\{
  \frac{1}{2\sqrt{6}} \rho^2_0(S_{\frac{1}{2}},1\,1) 
+  \frac{1}{2\sqrt{6}} \rho^2_0(S_{\frac{1}{2}},1\,2) 
\right\}
\nonumber \\  &&\!\!\!
+ J^2_0(\nu_{P_{\frac{3}{2}},S_{\frac{1}{2}}})
\left\{
   \frac{1}{2\sqrt{6}} \rho^0_0(S_{\frac{1}{2}},1\,1) 
-  \frac{1}{4\sqrt{3}} \rho^2_0(S_{\frac{1}{2}},1\,1) 
+  \frac{1}{4\sqrt{3}} \rho^2_0(S_{\frac{1}{2}},1\,2) 
\right\}
\Bigg]
\end{eqnarray}
\begin{eqnarray}
\dot{\rho}^0_0(P_{\frac{3}{2}},1\,1)
\!\!\!&=&\!\!\!
-\Big[
   A(P_{\frac{3}{2}}  \!\!\rightarrow\!\!  S_{\frac{1}{2}}) 
+  D^0(P_{\frac{3}{2}},1\,1) 
+  D^0(P_{\frac{3}{2}},1\,1 \!\!\rightarrow\!\! P_{\frac{1}{2}},1\,1)
\Big]  \rho_0^0(P_{\frac{3}{2}},1\,1) 
\\ &&\!\!\!
+ B(S_{\frac{1}{2}} \!\!\rightarrow\!\! P_{\frac{3}{2}})
\Bigg[
J_0^0(\nu_{P_{\frac{3}{2}},S_{\frac{1}{2}}})
\left\{
   \frac{5}{12} \rho_0^0(S_{\frac{1}{2}},1\,1) 
+  \frac{1}{4 \sqrt{15}}  \rho_0^0(S_{\frac{1}{2}},2\,2) 
\right\}
\nonumber \\&&\!\!\!
+J_0^2(\nu_{P_{\frac{3}{2}},S_{\frac{1}{2}}})
\left\{
-   \frac{5}{24}  \rho_0^2(S_{\frac{1}{2}},1\,1) 
+   \frac{1}{8}  \rho_0^2(S_{\frac{1}{2}},1\,2)
-  \frac{1}{8} \rho_0^2(S_{\frac{1}{2}},2\,1) 
+  \frac{1}{40} \sqrt{\frac{7}{3}} \rho_0^2(S_{\frac{1}{2}},2\,2)  
\right\}
\Bigg]
\nonumber \\&&\!\!\!
+  D^0(P_{\frac{1}{2}},1\,1 \!\!\rightarrow\!\! P_{\frac{3}{2}},1\,1) \rho_0^0(P_{\frac{1}{2}},1\,1)
+  D^0(P_{\frac{1}{2}},2\,2 \!\!\rightarrow\!\! P_{\frac{3}{2}},1\,1) \rho_0^0(P_{\frac{1}{2}},2\,2)
+  D^0(P_{\frac{3}{2}} 0\,0 \!\!\rightarrow\!\! P_{\frac{3}{2}},1\,1) \rho_0^0(P_{\frac{3}{2}},0\,0)
\nonumber  \\ &&\!\!\!
+  D^0(P_{\frac{3}{2}},2\,2 \!\!\rightarrow\!\! P_{\frac{3}{2}},1\,1) \rho_0^0(P_{\frac{3}{2}},2\,2)
+  D^0(P_{\frac{3}{2}},3\,3 \!\!\rightarrow\!\! P_{\frac{3}{2}},1\,1) \rho_0^0(P_{\frac{3}{2}},3\,3)
\nonumber 
\end{eqnarray}
\begin{eqnarray}
\dot{\rho}^2_0(P_{\frac{3}{2}},1\,1)
\!\!\!&=&\!\!\!
-\Big[
   A(P_{\frac{3}{2}}  \!\!\rightarrow\!\!  S_{\frac{1}{2}}) 
+  D^2(P_{\frac{3}{2}},1\,1)
+  D^0(P_{\frac{3}{2}},1\,1 \!\!\rightarrow\!\! P_{\frac{1}{2}},1\,1) 
\Big]  \rho_0^2(P_{\frac{3}{2}},1\,1)  
\\&& \!\!\!
+ B(S_{\frac{1}{2}} \!\!\rightarrow\!\! P_{\frac{3}{2}})
\Bigg[
J_0^0(\nu_{P_{\frac{3}{2}},S_{\frac{1}{2}}})
\left\{
-  \frac{5}{24}  \rho_0^2(S_{\frac{1}{2}},1\,1) 
-  \frac{1}{8}  \rho_0^2(S_{\frac{1}{2}},1\,2) 
+  \frac{1}{8}  \rho_0^2(S_{\frac{1}{2}},2\,1) 
+  \frac{1}{40} \sqrt{\frac{7}{3}} \rho_0^2(S_{\frac{1}{2}},2\,2) 
\right\}
\nonumber  \\&&\!\!\!
+ J_0^2(\nu_{P_{\frac{3}{2}},S_{\frac{1}{2}}})
\left\{
-  \frac{5}{24}  \rho_0^0(S_{\frac{1}{2}},1\,1) 
+  \frac{1}{40 \sqrt{15}}  \rho_0^0(S_{\frac{1}{2}},2\,2) 
-  \frac{5}{12 \sqrt{2}}   \rho_0^2(S_{\frac{1}{2}},1\,1) 
+  \frac{1}{20 \sqrt{42}}  \rho_0^2(S_{\frac{1}{2}},2\,2) 
\right.
\nonumber  \\&&\!\!\!
\left.
+  \frac{3}{10} \sqrt{\frac{3}{70}}  \rho_0^4(S_{\frac{1}{2}},2\,2) 
\right\}
\Bigg]
+  D^2(P_{\frac{1}{2}},1\,1 \!\!\rightarrow\!\! P_{\frac{3}{2}},1\,1) \rho_0^2(P_{\frac{1}{2}},1\,1)  
+  D^2(P_{\frac{1}{2}},2\,2 \!\!\rightarrow\!\! P_{\frac{3}{2}},1\,1) \rho_0^2(P_{\frac{1}{2}},2\,2)
\nonumber  \\&&\!\!\!
+  D^2(P_{\frac{3}{2}},2\,2 \!\!\rightarrow\!\! P_{\frac{3}{2}},1\,1) \rho_0^2(P_{\frac{3}{2}},2\,2)
+  D^2(P_{\frac{3}{2}},3\,3 \!\!\rightarrow\!\! P_{\frac{3}{2}},1\,1) \rho_0^2(P_{\frac{3}{2}},3\,3)
\nonumber
\end{eqnarray}
\begin{eqnarray}
\dot{\rho}^2_0(P_{\frac{3}{2}},1\,2)
\!\!\!&=&\!\!\!
- \Big[
   A(P_{\frac{3}{2}}  \!\!\rightarrow\!\!  S_{\frac{1}{2}}) 
+  D^2(P_{\frac{3}{2}},1\,2)
+  D^0(P_{\frac{3}{2}},1\,2 \!\!\rightarrow\!\! P_{\frac{1}{2}},1\,2)  
  \Big]   \rho_0^2(P_{\frac{3}{2}},1\,2) 
 \\ &&\!\!\!
+ B(S_{\frac{1}{2}} \!\!\rightarrow\!\! P_{\frac{3}{2}})
\Bigg[
J_0^0(\nu_{P_{\frac{3}{2}},S_{\frac{1}{2}}})
\left\{
    \frac{\sqrt{5}}{8}  \rho_2^2(S_{\frac{1}{2}},1\,1) 
+    \frac{\sqrt{5}}{24}   \rho_0^2(S_{\frac{1}{2}},1\,2)  
+    \frac{1}{8\sqrt{5}}    \rho_0^2(S_{\frac{1}{2}},2\,1)
+    \frac{1}{8} \sqrt{\frac{7}{15}}  \rho_0^2(S_{\frac{1}{2}},2\,2)  
\right\}
\nonumber  \\&&\!\!\!
+ J_0^2(\nu_{P_{\frac{3}{2}},S_{\frac{1}{2}}})
\left\{
-  \frac{\sqrt{5}}{8}    \rho_0^0(S_{\frac{1}{2}},1\,1) 
+  \frac{ \sqrt{3}}{40}  \rho_0^0(S_{\frac{1}{2}},2\,2)
-   \frac{1}{12} \sqrt{\frac{5}{2}} \rho_0^2(S_{\frac{1}{2}},1\,2) 
-   \frac{1}{4 \sqrt{10}}  \rho_0^2(S_{\frac{1}{2}},2\,1) 
\right.
\nonumber  \\&&\!\!\!
\left.
+   \frac{1}{2 \sqrt{210}}   \rho_0^2(S_{\frac{1}{2}},2\,2) 
-   \frac{1}{10} \sqrt{\frac{3}{14}}  \rho_0^4(S_{\frac{1}{2}},2\,2) 
\right\}
\Bigg]
 \nonumber
\end{eqnarray}
\begin{eqnarray}
\dot{\rho}^2_0(P_{\frac{3}{2}},1\,3)
\!\!\!&=&\!\!\!
-  \Big[
   A(P_{\frac{3}{2}}  \!\!\rightarrow\!\!  S_{\frac{1}{2}}) 
+  D^2(P_{\frac{3}{2}},1\,3) 
   \Big]  \rho_0^2(P_{\frac{3}{2}},1\,3)  
+  B(S_{\frac{1}{2}} \!\!\rightarrow\!\! P_{\frac{3}{2}})
\Bigg[
J_0^0(\nu_{P_{\frac{3}{2}},S_{\frac{1}{2}}})
\left\{
  \frac{1}{6} \sqrt{\frac{7}{2}} \rho_0^2(S_{\frac{1}{2}},1\,2)   
\right.
  \\&&\!\!\!
\left.
+ \frac{1}{10 \sqrt{6}}    \rho_0^2(S_{\frac{1}{2}},2\,2) 
\right\}
\nonumber   \\&&\!\!\!
+  J_0^2(\nu_{P_{\frac{3}{2}},S_{\frac{1}{2}}})
\left\{
    \frac{1}{10} \sqrt{\frac{21}{10}} \rho_0^0(S_{\frac{1}{2}},2\,2) 
-   \frac{5}{12\sqrt{7}}  \rho_0^2(S_{\frac{1}{2}},1\,2) 
-   \frac{23}{140 \sqrt{3}} \rho_0^2(S_{\frac{1}{2}},2\,2) 
+    \frac{1}{70}\sqrt{\frac{3}{5}}  \rho_0^4(S_{\frac{1}{2}},2\,2) 
\right\}
\Bigg]
 \nonumber 
\end{eqnarray}
\begin{eqnarray}
\dot{\rho}^4_0(P_{\frac{3}{2}},1\,3)
\!\!\!&=&\!\!\!
-  \Big[
   A(P_{\frac{3}{2}}  \!\!\rightarrow\!\!  S_{\frac{1}{2}}) 
+  D^4(P_{\frac{3}{2}},1\,3) 
   \Big]  \rho_0^4(P_{\frac{3}{2}},1\,3)
+ B(S_{\frac{1}{2}} \!\!\rightarrow\!\! P_{\frac{3}{2}})
\Bigg[
   \frac{1}{2 \sqrt{10}} J_0^0(\nu_{P_{\frac{3}{2}},S_{\frac{1}{2}}})  \rho_0^4(S_{\frac{1}{2}},2\,2) 
\\ &&\!\!\!
+  J_0^2(\nu_{P_{\frac{3}{2}},S_{\frac{1}{2}}})
\left\{
-  \frac{1}{2} \sqrt{\frac{3}{7}}  \rho_0^2(S_{\frac{1}{2}},1\,2) 
+  \frac{3}{70}  \rho_0^2(S_{\frac{1}{2}},2\,2)   
-  \frac{\sqrt{5}}{28}  \rho_0^4(S_{\frac{1}{2}},2\,2)  
\right\}
\Bigg]
\nonumber
\end{eqnarray}
\begin{eqnarray}
\dot{\rho}^2_0(P_{\frac{3}{2}},2\,0)
\!\!\!&=&\!\!\!
-  \Big[
   A(P_{\frac{3}{2}}  \!\!\rightarrow\!\!  S_{\frac{1}{2}}) 
+  D^2(P_{\frac{3}{2}},2\,0) 
   \Big]  \rho_0^2(P_{\frac{3}{2}},2\,0)
+ B(S_{\frac{1}{2}} \!\!\rightarrow\!\! P_{\frac{3}{2}})  
\Bigg[
J_0^0(\nu_{P_{\frac{3}{2}},S_{\frac{1}{2}}})
\left\{
   \frac{1}{2\sqrt{6}} \rho_0^2(S_{\frac{1}{2}},1\,1) 
\right.     
\\&&\!\!\!
\left.
-  \frac{1}{2\sqrt{6}} \rho_0^2(S_{\frac{1}{2}},2\,1)  
\right\}
+ J_0^2(\nu_{P_{\frac{3}{2}},S_{\frac{1}{2}}})
\left\{
 \frac{1}{2\sqrt{6}} \rho_0^0(S_{\frac{1}{2}},1\,1) 
   -\frac{1}{4\sqrt{3}} \rho_0^2(S_{\frac{1}{2}},1\,1)  
   -\frac{1}{4\sqrt{3}} \rho_0^2(S_{\frac{1}{2}},2\,1) 
\right\}
\Bigg]
  \nonumber 
\end{eqnarray}
\begin{eqnarray}
\dot{\rho}^2_0(P_{\frac{3}{2}},2\,1)
\!\!\!&=&\!\!\!
-  \Big[
   A(P_{\frac{3}{2}}  \!\!\rightarrow\!\!  S_{\frac{1}{2}}) 
+  D^2(P_{\frac{3}{2}},2\,1) 
+  D^0(P_{\frac{3}{2}},2\,1 \!\!\rightarrow\!\! P_{\frac{1}{2}},2\,1)
   \Big]  \rho_0^2(P_{\frac{3}{2}},2\,1) 
\\&&\!\!\!
+  B(S_{\frac{1}{2}} \!\!\rightarrow\!\! P_{\frac{3}{2}})
\Bigg[
J_0^0(\nu_{P_{\frac{3}{2}},S_{\frac{1}{2}}})
\left\{
-   \frac{\sqrt{5}}{8}   \rho_0^2(S_{\frac{1}{2}},1\,1) 
+   \frac{1}{8\sqrt{5}}  \rho_0^2(S_{\frac{1}{2}},1\,2)
+   \frac{\sqrt{5}}{24}  \rho_0^2(S_{\frac{1}{2}},2\,1)
-   \frac{1}{8} \sqrt{\frac{7}{15}}  \rho_0^2(S_{\frac{1}{2}},2\,2)   
\right\}
\nonumber  \\&&\!\!\!
+  J_0^2(\nu_{P_{\frac{3}{2}},S_{\frac{1}{2}}})
\left\{  
   \frac{\sqrt{5}}{8}   \rho_0^0(S_{\frac{1}{2}},1\,1) 
-  \frac{\sqrt{3}}{40}   \rho_0^0(S_{\frac{1}{2}},2\,2) 
-  \frac{1}{4 \sqrt{10}} \rho_0^2(S_{\frac{1}{2}},1\,2) 
-   \frac{1}{12} \sqrt{\frac{5}{2}} \rho_0^2(S_{\frac{1}{2}},2\,1) .
\right.
\nonumber  \\&&\!\!\!
\left.
-   \frac{1}{2 \sqrt{210}} \rho_0^2(S_{\frac{1}{2}},2\,2) 
+   \frac{1}{10} \sqrt{\frac{3}{14}}  \rho_0^4(S_{\frac{1}{2}},2\,2) 
\right\}
\Bigg]
\nonumber  
\end{eqnarray}
\begin{eqnarray}
\dot{\rho}^0_0(P_{\frac{3}{2}},2\,2)
\!\!\!&=&\!\!\!
-  \Big[
   A(P_{\frac{3}{2}}  \!\!\rightarrow\!\!  S_{\frac{1}{2}}) 
+  D^0(P_{\frac{3}{2}},2\,2)
+  D^0(P_{\frac{3}{2}},2\,2 \!\!\rightarrow\!\! P_{\frac{1}{2}},2\,2)
   \Big]  \rho_0^0(P_{\frac{3}{2}},2\,2)
\\&&\!\!\!
+  B(S_{\frac{1}{2}} \!\!\rightarrow\!\! P_{\frac{3}{2}})
\Bigg[
J_0^0(\nu_{P_{\frac{3}{2}},S_{\frac{1}{2}}})
\left\{
\frac{1}{4} \sqrt{\frac{5}{3}} \rho_0^0(S_{\frac{1}{2}},1\,1) 
+\frac{1}{4} \rho_0^0(S_{\frac{1}{2}},2\,2) 
\right\}
\nonumber  \\&&\!\!\!
+  J_0^2(\nu_{P_{\frac{3}{2}},S_{\frac{1}{2}}})
\left\{
   \frac{1}{8 \sqrt{15}} \rho_0^2(S_{\frac{1}{2}},1\,1) 
-  \frac{1}{8} \sqrt{\frac{3}{5}}  \rho_0^2(S_{\frac{1}{2}},1\,2)  
+  \frac{1}{8} \sqrt{\frac{3}{5}} \rho_0^2(S_{\frac{1}{2}},2\,1) 
-  \frac{1}{8} \sqrt{\frac{7}{5}} \rho_0^2(S_{\frac{1}{2}},2\,2) 
\right\}
\Bigg]
\nonumber  \\&&\!\!\!
+  D^0(P_{\frac{1}{2}},1\,1,P_{\frac{3}{2}},2\,2) \rho_0^0(P_{\frac{1}{2}},1\,1) 
+  D^0(P_{\frac{1}{2}},2\,2 \!\!\rightarrow\!\! P_{\frac{3}{2}},2\,2)\rho_0^0(P_{\frac{1}{2}},2\,2)
+  D^0(P_{\frac{3}{2}},0\,0 \!\!\rightarrow\!\! P_{\frac{3}{2}},2\,2)\rho_0^0(P_{\frac{3}{2}},0\,0)
\nonumber  \\ &&\!\!\!
+  D^0(P_{\frac{3}{2}},1 1 \!\!\rightarrow\!\! P_{\frac{3}{2}},2\,2) \rho_0^0(P_{\frac{3}{2}},1\,1) 
+  D^0(P_{\frac{3}{2}},3\,3 \!\!\rightarrow\!\! P_{\frac{3}{2}},2\,2) \rho_0^0(P_{\frac{3}{2}},3\,3)
\nonumber
\end{eqnarray}
\begin{eqnarray}
\dot{\rho }^2_0(P_{\frac{3}{2}},2\,2)
\!\!\!&=&\!\!\!
-  \Big[
   A(P_{\frac{3}{2}}  \!\!\rightarrow\!\!  S_{\frac{1}{2}}) 
+  D^2(P_{\frac{3}{2}},2\,2) 
+  D^0(P_{\frac{3}{2}},2\,2 \!\!\rightarrow\!\! P_{\frac{1}{2}},2\,2)
   \Big] \rho_0^2(P_{\frac{3}{2}},2\,2)
\\&&\!\!\!
+  B(S_{\frac{1}{2}} \!\!\rightarrow\!\! P_{\frac{3}{2}})
\Bigg[
J_0^0(\nu_{P_{\frac{3}{2}},S_{\frac{1}{2}}})
\left\{
   \frac{1}{8} \sqrt{\frac{7}{3}} \rho_0^2(S_{\frac{1}{2}},1\,1) 
-  \frac{1}{8} \sqrt{\frac{7}{3}}  \rho_0^2(S_{\frac{1}{2}},1\,2) 
+  \frac{1}{8} \sqrt{\frac{7}{3}}  \rho_0^2(S_{\frac{1}{2}},2\,1)
+  \frac{1}{8} \rho_0^2(S_{\frac{1}{2}},2\,2) 
\right\}
\nonumber \\&&\!\!\!
+   J_0^2(\nu_{P_{\frac{3}{2}},S_{\frac{1}{2}}})
\left\{
\frac{1}{8} \sqrt{\frac{7}{3}} \rho_0^0(S_{\frac{1}{2}},1\,1) 
-  \frac{1}{8} \sqrt{\frac{7}{5}} \rho_0^0(S_{\frac{1}{2}},2\,2) 
+  \frac{1}{4 \sqrt{42}} \rho_0^2(S_{\frac{1}{2}},1\,1) 
-  \frac{1}{2 \sqrt{42}} \rho_0^2(S_{\frac{1}{2}},1\,2)
\right.
\nonumber \\&&\!\!\!
\left.
+   \frac{1}{2 \sqrt{42}} \rho_0^2(S_{\frac{1}{2}},2\,1) 
-  \frac{5}{28 \sqrt{2}} \rho_0^2(S_{\frac{1}{2}},2\,2)
-  \frac{3}{14\sqrt{10}}  \rho_0^4(S_{\frac{1}{2}},2\,2) 
\right\}
\Bigg]
\nonumber  \\&&\!\!\!
+  D^2(P_{\frac{1}{2}},1\,1  \!\!\rightarrow\!\!  P_{\frac{3}{2}},2\,2) \rho_0^2(P_{\frac{1}{2}},1\,1)
+  D^2(P_{\frac{1}{2}},2\,2  \!\!\rightarrow\!\!  P_{\frac{3}{2}},2\,2) \rho_0^2(P_{\frac{1}{2}},2\,2)
+  D^2(P_{\frac{3}{2}},1\,1  \!\!\rightarrow\!\!  P_{\frac{3}{2}},2\,2) \rho_0^2(P_{\frac{3}{2}},1\,1)
\nonumber  \\ &&\!\!\!
+  D^2(P_{\frac{3}{2}},3\,3  \!\!\rightarrow\!\!  P_{\frac{3}{2}},2\,2) \rho_0^2(P_{\frac{3}{2}},3\,3)
\nonumber
\end{eqnarray}
\begin{eqnarray}
\dot{\rho}^4_0(P_{\frac{3}{2}},2\,2)
\!\!\!&=&\!\!\!
-  \Big[
   A(P_{\frac{3}{2}}  \!\!\rightarrow\!\!  S_{\frac{1}{2}}) 
+  D^4(P_{\frac{3}{2}},2\,2) 
+  D^0(P_{\frac{3}{2}},2\,2 \!\!\rightarrow\!\! P_{\frac{1}{2}},2\,2)
   \Big]  \rho_0^4(P_{\frac{3}{2}},2\,2)
\\&&\!\!\!
+  B(S_{\frac{1}{2}} \!\!\rightarrow\!\! P_{\frac{3}{2}})
\Bigg[
-   \frac{1}{6} J_0^0(\nu_{P_{\frac{3}{2}},S_{\frac{1}{2}}})   \rho_0^4(S_{\frac{1}{2}},2\,2) 
+   J_0^2(\nu_{P_{\frac{3}{2}},S_{\frac{1}{2}}})   
\left\{
   \frac{3}{2} \sqrt{\frac{3}{70}}  \rho_0^2(S_{\frac{1}{2}},1\,1) 
+  \frac{1}{2} \sqrt{\frac{3}{70}}  \rho_0^2(S_{\frac{1}{2}},1\,2) 
\right.
\nonumber  \\&&\!\!\!
\left.
-  \frac{1}{2} \sqrt{\frac{3}{70}}  \rho_0^2(S_{\frac{1}{2}},2\,1) 
-  \frac{3}{14 \sqrt{10}}           \rho_0^2(S_{\frac{1}{2}},2\,2) 
-  \frac{5}{21\sqrt{2}}             \rho_0^4(S_{\frac{1}{2}},2\,2)  
\right\}
\Bigg]
\nonumber  \\&&\!\!\!
+  D^4(P_{\frac{1}{2}},2\,2  \!\!\rightarrow\!\!  P_{\frac{3}{2}},2\,2) \rho_0^4(P_{\frac{1}{2}},2\,2)
+  D^4(P_{\frac{3}{2}},3\,3  \!\!\rightarrow\!\!  P_{\frac{3}{2}},2\,2) \rho_0^4(P_{\frac{3}{2}},3\,3) 
\nonumber
\end{eqnarray}
\begin{eqnarray}
\dot{\rho}^2_0(P_{\frac{3}{2}},2\,3)
\!\!\!&=&\!\!\!
-  \Big[
   A(P_{\frac{3}{2}}  \!\!\rightarrow\!\!  S_{\frac{1}{2}}) 
+  D^2(P_{\frac{3}{2}},2\,3) 
   \Big]   \rho_0^2(P_{\frac{3}{2}},2\,3)
+   B(S_{\frac{1}{2}} \!\!\rightarrow\!\! P_{\frac{3}{2}})
\Bigg[
    J_0^0(\nu_{P_{\frac{3}{2}},S_{\frac{1}{2}}})
\left\{
   \frac{1}{2} \sqrt{\frac{7}{15}}  \rho_0^2(S_{\frac{1}{2}},1\,2)
\right.    
\\&&\!\!\!
\left.
+  \frac{1}{2\sqrt{5}}              \rho_0^2(S_{\frac{1}{2}},2\,2) 
\right\}
\nonumber  \\&&\!\!\!
+   J_0^2(\nu_{P_{\frac{3}{2}},S_{\frac{1}{2}}})
\left\{
-  \frac{1}{10} \sqrt{7}    \rho_0^0(S_{\frac{1}{2}},2\,2)
+  \frac{1}{2\sqrt{210}}    \rho_0^2(S_{\frac{1}{2}},1\,2) 
-   \frac{1}{14 \sqrt{10}}  \rho_0^2(S_{\frac{1}{2}},2\,2) 
+   \frac{3}{70\sqrt{2}}    \rho_0^4(S_{\frac{1}{2}},2\,2)
\right\}
\Bigg]
\nonumber
\end{eqnarray}
\begin{eqnarray}
\dot{\rho}^4_0(P_{\frac{3}{2}},2\,3)
\!\!\!&=&\!\!\!
-  \Big[
   A(P_{\frac{3}{2}}  \!\!\rightarrow\!\!  S_{\frac{1}{2}}) 
+  D^4(P_{\frac{3}{2}},2\,3) 
   \Big]  \rho_0^4(P_{\frac{3}{2}},2\,3)
+  B(S_{\frac{1}{2}} \!\!\rightarrow\!\! P_{\frac{3}{2}})
\Bigg[
   \frac{1}{6} \sqrt{\frac{5}{2}} J_0^0(\nu_{P_{\frac{3}{2}},S_{\frac{1}{2}}})  \rho_0^4(S_{\frac{1}{2}},2\,2) 
\\&&\!\!\!
+  J_0^2(\nu_{P_{\frac{3}{2}},S_{\frac{1}{2}}})
\left\{
   \frac{1}{2} \sqrt{\frac{3}{7}} \rho_0^2(S_{\frac{1}{2}},1\,2) 
-  \frac{3}{14} \rho_0^2(S_{\frac{1}{2}},2\,2)   
+  \frac{\sqrt{5}}{84}  \rho_0^4(S_{\frac{1}{2}},2\,2) 
\right\}
\Bigg]
\nonumber
\end{eqnarray}
\begin{eqnarray}
\dot{\rho}^2_0(P_{\frac{3}{2}},3\,1)
\!\!\!&=&\!\!\!
-  \Big[
   A(P_{\frac{3}{2}}  \!\!\rightarrow\!\!  S_{\frac{1}{2}}) 
+  D^2(P_{\frac{3}{2}},3\,1) 
   \Big]   \rho_0^2(P_{\frac{3}{2}},3\,1)
+  B(S_{\frac{1}{2}} \!\!\rightarrow\!\! P_{\frac{3}{2}})
\Bigg[
   J_0^0(\nu_{P_{\frac{3}{2}},S_{\frac{1}{2}}})
\left\{
-\frac{1}{6} \sqrt{\frac{7}{2}}   \rho_0^2(S_{\frac{1}{2}},2\,1)
\right.    
\\&&\!\!\!
\left.
+\frac{1}{10 \sqrt{6}} \rho_0^2(S_{\frac{1}{2}},2\,2) 
\right\} 
\nonumber  \\&&\!\!\! 
+    J_0^2(\nu_{P_{\frac{3}{2}},S_{\frac{1}{2}}})
\left\{
\frac{1}{10} \sqrt{\frac{21}{10}} \rho_0^0(S_{\frac{1}{2}},2\,2)
+\frac{5}{12\sqrt{7}} \rho_0^2(S_{\frac{1}{2}},2\,1) 
-\frac{23}{140 \sqrt{3}} \rho_0^2(S_{\frac{1}{2}},2\,2) 
+\frac{1}{70} \sqrt{\frac{3}{5}}  \rho_0^4(S_{\frac{1}{2}},2\,2) 
\right\}
\Bigg]  
 \nonumber  
\end{eqnarray}
\begin{eqnarray}
\dot{\rho}^4_0(P_{\frac{3}{2}},3\,1)
\!\!\!&=&\!\!\!
-  \Big[
   A(P_{\frac{3}{2}}  \!\!\rightarrow\!\!  S_{\frac{1}{2}}) 
+  D^4(P_{\frac{3}{2}},3\,1) 
   \Big] \rho_0^4(P_{\frac{3}{2}},3\,1)
+   B(S_{\frac{1}{2}} \!\!\rightarrow\!\! P_{\frac{3}{2}})
\Bigg[
\frac{1}{2 \sqrt{10}}  J_0^0(\nu_{P_{\frac{3}{2}},S_{\frac{1}{2}}}) \rho_0^4(S_{\frac{1}{2}},2\,2)
\\&& \!\!\!
+   J_0^2(\nu_{P_{\frac{3}{2}},S_{\frac{1}{2}}})
\left\{
 \frac{1}{2} \sqrt{\frac{3}{7}}  \rho_0^2(S_{\frac{1}{2}},2\,1) 
+\frac{3}{70}  \rho_0^2(S_{\frac{1}{2}},2\,2)      
-\frac{1}{28} \sqrt{5} \rho_0^4(S_{\frac{1}{2}},2\,2)  
\right\}
\Bigg]
\nonumber
\end{eqnarray}
\begin{eqnarray}
\dot{\rho}^2_0(P_{\frac{3}{2}},3\,2)
\!\!\!&=&\!\!\!
- \Big[
  A(P_{\frac{3}{2}}  \!\!\rightarrow\!\!  S_{\frac{1}{2}}) 
+ D^2(P_{\frac{3}{2}},3\,2) 
  \Big]  \rho_0^2(P_{\frac{3}{2}},3\,2)
+
B(S_{\frac{1}{2}} \!\!\rightarrow\!\! P_{\frac{3}{2}})
\Bigg[ 
   J_0^0(\nu_{P_{\frac{3}{2}},S_{\frac{1}{2}}})
\left\{
  \frac{1}{2} \sqrt{\frac{7}{15}}    \rho_0^2(S_{\frac{1}{2}},2\,1) 
\right.
\\&&\!\!\!
\left.
-\frac{1}{2\sqrt{5}}   \rho_0^2(S_{\frac{1}{2}},2\,2) 
\right\}
\nonumber \\&&\!\!\!
+  J_0^2(\nu_{P_{\frac{3}{2}},S_{\frac{1}{2}}}) 
\left\{
\frac{1}{2\sqrt{210}} \rho_0^2(S_{\frac{1}{2}},2\,1) 
+\frac{1}{10} \sqrt{7}  \rho_0^0(S_{\frac{1}{2}},2\,2) 
+\frac{1}{14 \sqrt{10}} \rho_0^2(S_{\frac{1}{2}},2\,2) 
-\frac{3}{70\sqrt{2}}   \rho_0^4(S_{\frac{1}{2}},2\,2) 
\right\}
\Bigg]
\nonumber
\end{eqnarray}
\begin{eqnarray}
\dot{\rho}^4_0(P_{\frac{3}{2}},3\,2)
\!\!\!&=&\!\!\!
-  \Big[
   A(P_{\frac{3}{2}}  \!\!\rightarrow\!\!  S_{\frac{1}{2}})
+  D^4(P_{\frac{3}{2}},3\,2) 
   \Big]    \rho_0^4(P_{\frac{3}{2}},3\,2)
+  B(S_{\frac{1}{2}} \!\!\rightarrow\!\! P_{\frac{3}{2}})
\Bigg[
-  \frac{1}{6} \sqrt{\frac{5}{2}} J_0^0(\nu_{P_{\frac{3}{2}},S_{\frac{1}{2}}}) \rho_0^4(S_{\frac{1}{2}},2\,2) 
\\&&\!\!\!
+  J_0^2(\nu_{P_{\frac{3}{2}},S_{\frac{1}{2}}})
\left\{
\frac{1}{2} \sqrt{\frac{3}{7}}  \rho_0^2(S_{\frac{1}{2}},2\,1) 
+\frac{3}{14}  \rho_0^2(S_{\frac{1}{2}},2\,2)     
-\frac{\sqrt{5}}{84}   \rho_0^4(S_{\frac{1}{2}},2\,2)  
\right\}
\Bigg]
 \nonumber
\end{eqnarray}
\begin{eqnarray}
\dot{\rho}^0_0(P_{\frac{3}{2}},3\,3)
\!\!\!&=&\!\!\!
-  \Big[
   A(P_{\frac{3}{2}}  \!\!\rightarrow\!\!  S_{\frac{1}{2}}) 
+  D^0(P_{\frac{3}{2}},3\,3) 
   \Big]   \rho_0^0(P_{\frac{3}{2}},3\,3)
+  B(S_{\frac{1}{2}} \!\!\rightarrow\!\! P_{\frac{3}{2}})
\Bigg[
\frac{1}{2} \sqrt{\frac{7}{5}}  J_0^0(\nu_{P_{\frac{3}{2}},S_{\frac{1}{2}}})  \rho_0^0(S_{\frac{1}{2}},2\,2) 
\\&&\!\!\!
+ \frac{1}{10} J_0^2(\nu_{P_{\frac{3}{2}},S_{\frac{1}{2}}})     \rho_0^2(S_{\frac{1}{2}},2\,2)
\Bigg] 
+  D^0(P_{\frac{1}{2}},1\,1,P_{\frac{3}{2}},3\,3)  \rho_0^0(P_{\frac{1}{2}},1\,1)
+  D^0(P_{\frac{1}{2}},2\,2,P_{\frac{3}{2}},3\,3)  \rho_0^0(P_{\frac{1}{2}},2\,2)
\nonumber  \\&&\!\!\!
+  D^0(P_{\frac{3}{2}},0\,0  \!\!\rightarrow\!\!  P_{\frac{3}{2}},3\,3)  \rho_0^0(P_{\frac{3}{2}},0\,0)
+  D^0(P_{\frac{3}{2}},1\,1  \!\!\rightarrow\!\!  P_{\frac{3}{2}},3\,3)   \rho_0^0(P_{\frac{3}{2}},1\,1)
+  D^0(P_{\frac{3}{2}},2\,2  \!\!\rightarrow\!\!  P_{\frac{3}{2}},3\,3)  \rho_0^0(P_{\frac{3}{2}},2\,2)
  \nonumber 
\end{eqnarray}
\begin{eqnarray}
\dot{\rho}^2_0(P_{\frac{3}{2}},3\,3)
\!\!\!&=&\!\!\!
-  \Big[
   A(P_{\frac{3}{2}}  \!\!\rightarrow\!\!  S_{\frac{1}{2}}) 
+  D^2(P_{\frac{3}{2}},3\,3) 
   \Big]   \rho_0^2(P_{\frac{3}{2}},3\,3)
+  B(S_{\frac{1}{2}} \!\!\rightarrow\!\! P_{\frac{3}{2}})
\Bigg[
   \frac{\sqrt{6}}{5}   J_0^0(\nu_{P_{\frac{3}{2}},S_{\frac{1}{2}}})  \rho_0^2(S_{\frac{1}{2}},2\,2)    
\\&&\!\!\!
+  J_0^2(\nu_{P_{\frac{3}{2}},S_{\frac{1}{2}}})
\left\{
\frac{1}{5} \sqrt{\frac{21}{10}} \rho_0^0(S_{\frac{1}{2}},2\,2) 
+\frac{2\sqrt{3}}{35}  \rho_0^2(S_{\frac{1}{2}},2\,2) 
+\frac{1}{35} \sqrt{\frac{3}{5}} \rho_0^4(S_{\frac{1}{2}},2\,2) 
\right\}
\nonumber  \\&&\!\!\!
+  D^2(P_{\frac{1}{2}},1\,1  \!\!\rightarrow\!\!  P_{\frac{3}{2}},3\,3)  \rho_0^2(P_{\frac{1}{2}},1\,1)
+  D^2(P_{\frac{1}{2}},2\,2  \!\!\rightarrow\!\!  P_{\frac{3}{2}},3\,3)   \rho_0^2(P_{\frac{1}{2}},2\,2)
+ D^2(P_{\frac{3}{2}},1\,1  \!\!\rightarrow\!\!  P_{\frac{3}{2}},3\,3)  \rho_0^2(P_{\frac{3}{2}},1\,1)
\nonumber  \\&&\!\!\!
+D^2(P_{\frac{3}{2}},2\,2  \!\!\rightarrow\!\!  P_{\frac{3}{2}},3\,3)  \rho_0^2(P_{\frac{3}{2}},2\,2)
  \nonumber 
\end{eqnarray}
\begin{eqnarray}
\dot{\rho}^4_0(P_{\frac{3}{2}},3\,3)
\!\!\!&=&\!\!\!
-  \Big[
   A(P_{\frac{3}{2}}  \!\!\rightarrow\!\!  S_{\frac{1}{2}}) 
+  D^4(P_{\frac{3}{2}},3\,3) 
   \Big]   \rho_0^4(P_{\frac{3}{2}},3\,3)
+  B(S_{\frac{1}{2}} \!\!\rightarrow\!\! P_{\frac{3}{2}}) 
\Bigg[
   \frac{1}{6} \sqrt{\frac{11}{5}}  J_0^0(\nu_{P_{\frac{3}{2}},S_{\frac{1}{2}}})  \rho_0^4(S_{\frac{1}{2}},2\,2) 
\\ &&\!\!\!
+   J_0^2(\nu_{P_{\frac{3}{2}},S_{\frac{1}{2}}})
\left\{
   \frac{3\sqrt{22}}{35}    \rho_0^2(S_{\frac{1}{2}},2\,2) 
+  \frac{1}{21} \sqrt{\frac{10}{11}}   \rho_0^4(S_{\frac{1}{2}},2\,2) 
\right\}
 \Bigg]
+  D^4(P_{\frac{1}{2}},2\,2  \!\!\rightarrow\!\!  P_{\frac{3}{2}},3\,3)  \rho_0^4(P_{\frac{1}{2}},2\,2)
\nonumber  \\ &&\!\!\!
+  D^4(P_{\frac{3}{2}},2\,2  \!\!\rightarrow\!\!  P_{\frac{3}{2}},3\,3)  \rho_0^4(P_{\frac{3}{2}},2\,2) 
\nonumber 
\end{eqnarray}
\begin{eqnarray}
\dot{\rho}^6_0(P_{\frac{3}{2}},3\,3)
\!\!\!&=&\!\!\!
- \Big[
   A(P_{\frac{3}{2}} \!\!\rightarrow\!\! S_{\frac{1}{2}}) 
+  D^6(P_{\frac{3}{2}},3\,3) 
  \Big]  \rho^6_0(P_{\frac{3}{2}},3\,3) 
+ \frac{1}{2} \sqrt{\frac{15}{11}}  B(S_{\frac{1}{2}} \!\!\rightarrow\!\! P_{\frac{3}{2}}) J_0^2(\nu_{P_{\frac{3}{2}},S_{\frac{1}{2}}})  \rho^4_0(S_{\frac{1}{2}},2\,2) 
\end{eqnarray}
\end{widetext}
\end{appendix}

\end{document}